\documentclass[sn-mathphys-num]{sn-jnl}

\usepackage{graphicx}%
\usepackage{multirow}%
\usepackage{amsmath,amssymb,amsfonts}%
\usepackage{amsthm}%
\usepackage{mathrsfs}%
\usepackage[title]{appendix}%
\usepackage{xcolor}%
\usepackage{textcomp}%
\usepackage{manyfoot}%
\usepackage{booktabs}%
\usepackage{algorithm}%
\usepackage{algorithmicx}%
\usepackage{algpseudocode}%
\usepackage{listings}%
\usepackage{braket}%

\DeclareMathOperator{\Tr}{Tr}
\usepackage{marvosym}

\theoremstyle{thmstyleone}%
\theoremstyle{thmstyletwo}%

\theoremstyle{thmstylethree}%

\begin{document}

\title[Quantum Computers, Quantum Computing and Quantum Thermodynamics]{Quantum Computers, Quantum Computing and Quantum Thermodynamics}

\author[  1,2]{\fnm{Fabrizio} \sur{Cleri}\footnote{on temporary leave at: LIMMS CNRS IRL2820, University of Tokyo, 4-6-1 Komaba, Meguro-Ku, Tokyo 153-8505, Japan. e-mail: fabrizio.cleri@univ-lille.fr}} 

\affil[1]{ \orgdiv{Institute of Electronics, Microelectronics and Nanotechnology}, \orgname{(IEMN CNRS UMR8520)}, \orgaddress{\street{Av. Poincar\'e}, \city{Villeneuve d'Ascq}, \postcode{59652}, \country{France}}}

\affil[2]{ \orgdiv{D\'epartement de Physique}, \orgname{Universit\'e de Lille}, \orgaddress{\city{Villeneuve d'Ascq}, \postcode{59650}, \country{France}}}



\abstract{ Quantum thermodynamics aims at extending standard thermodynamics
and non-equilibrium statistical physics to systems with sizes well below the thermodynamic limit.
A rapidly evolving research field, which promises to change
our understanding of the foundations of physics, while enabling the discovery of novel thermodynamic techniques and applications at the nanoscale.  Thermal management has turned into a major obstacle in pushing the limits of conventional digital computers, and could likely represent a crucial issue also for quantum computers.
The practical realization of quantum computers with superconducting loops requires working at cryogenic temperatures to eliminate thermal noise; ion-trap qubits need as well low temperatures to minimize collisional noise; in both cases, the sub-nanometric sizes also bring about thermal broadening of the quantum states; and even room-temperature photonic computers require cryogenic detectors. A number of thermal and thermodynamic questions therefore take center stage, such as quantum re-definitions of work and heat, thermalization and randomization of quantum states, the overlap of quantum and thermal fluctuations, and many other, even including a proper definition of temperature for the small open systems constantly out of equilibrium that are the qubits.
This overview provides an introductory perspective on a
selection of current trends in quantum thermodynamics and their impact on quantum computers and quantum computing, with a language accessible also to postgraduate students and researchers from different fields. }



\maketitle

\newpage
\tableofcontents

\section{Introduction}\label{sec1}

Quantum computing has gone a long way, from the dream of Richard Feynman expressed in his keynote paper of 1982,\cite{feynman82} to the increasingly sophisticated theoretical developments that followed in the '90s, to the first realizations of experimental quantum bits in the past 20 years.\cite{nature,china20,roadmap} Today we are just starting to see the first true quantum machines sporting some tens of qubits connected by fully reversible quantum gates, which attempt at solving theoretical benchmark challenges. It is not still the calculation of real-life problems, however they are getting a bit closer every few months, in a path that seems to echo the spectacular growth of digital microelectronics in the last half of the past century.

By looking at the latest developments among the main players in quantum hardware and software, Google's Quantum AI subsidiary first reported having reached “quantum advantage” in July 2019 with their Sycamore machine,\cite{sycam} a claim later challenged by IBM engineers. (see below) Both IBM and Google use qubits made with superconducting loops. IBM broke already the 100-wall with its 127-qubit Quantum Eagle,\cite{ibm21} and just announced its new milestone by the end of 2023 with his QuantumSystem-Two modular architecture including three Heron 133-qubit devices. Also Intel is engaged in both superconducting and spin qubit research: June 2023 unveiled its new TunnelFalls, a 12-qubit all-silicon chip. NVIDIA launched in March 2023 the DGX Quantum, a GPU-accelerated system, integrating their GraceHopper superchip with the OPX platform by Quantum Machines. Honeywell opted for trapped-ion qubits in their System-Model H1, 10-qubit first operational machine, already used for quantum chemistry simulations;\cite{quantin} a similar road to that followed by IonQ with their Aria 25-qubit machine. Notably, all these companies are making their computing platforms, or a scaled-down version thereof, publicly web-accessible to anybody for testing and running quantum codes via the internet. In June 2022, the US Senate passed the \textdollar250 billion Innovation and Competition Act, promoting quantum information technologies among the actions to ensure that the US semiconductor and information technology continue to play a leading role in the global economy. At the other shore of the Pacific, with the help of a multi-billion-dollar funding package and a \texteuro10 billion investment in a quantum information laboratory, China hopes to make significant breakthroughs in the field by 2030. Big names such as Alibaba and Baidu are engaged in sustaining R\&D (although Alibaba's quantum laboratory was suddenly shut down Nov. 2023). But already, one team at the University of Science and Technology of China in Hefei, reported achieving quantum advantage by using two radically different technologies, linear optics or superconducting qubits, just one year apart from each other.\cite{pan2020,pan2021}

\smallskip

As far as Europe, In October 2018 the European Commission launched the "Quantum Technologies Flagship" programme, to support hundreds of quantum science researchers over a ten-year period with a budget of \texteuro1 billion. The OpenSuperQPlus project  is a medium-term, 4-year project centered at the J\"ulich Research Center in Germany, assembling 28 partners from 10 EU countries. However, compared to the US and China market dominated by a few hi-tech giants, the EU panorama is richer in smaller partnerships and smaller companies.\cite{unicorns} UK, sadly no longer part of the European Union, has announced conspicuous investments, while initially relying on technology provided by the US start-up Rigetti, and other local solutions such as the OQC company in Oxford developing their "coaxmon".\cite{coax} As usual, other EU countries are proceeding working the two sides of the street, partly following EU guidelines and partly pushing national initiatives. Germany follows a strategy similar to UK, coupling \texteuro3 billion in national investment with US quantum technology imported from IBM. Netherlands launched its national quantum strategy in 2019 with \texteuro615 million and the Quantum Delta NL initiative to help quantum research and marketing in universities. France follows, as it often happens, a more original way, with a 5-year \texteuro1.8 billion funding initiative  (half of which coming from public money), the development of a large-scale quantum annealer (a somewhat different concept from the gate-based quantum computer) by the start-up company Pasqal,\cite{pasqal} and in parallel the Qandela \cite{qandela} just-announced photonic computer installed in the north of France. 

\smallskip

Then, why thermodynamics? Heat dissipation has always been a crucial problem for digital computers, and represents probably the biggest limit to a further expansion of the CMOS-based computing technology.\cite{kish2002,vala2018,bespa2022} Up to the '90s, the solution was to reduce the voltage levels, but now we are already at 0.7 V and this figure could not be reduced further. The heating problem has been exacerbated with the introduction of 3-dimensional design, that brought with it new issues of capacitive charging of the metal connections crossing in the vertical direction. The progressively reduced transistor dimension, now at limits reaching below the 10 nm, has the additional issue of self-heating because of the largely increased surface/volume ratio of the device. Switching is also at its limits: we have devices in our laboratories that can easily function at 100 GHz and more, however the fastest clock cycle adopted in real computing units cannot go above the 5-6 GHz, just because the rate of heat accumulation cannot be matched by a fast-enough rate of dissipation.
Quantum computers could be in principle even more sensitive to fluctuations and heat dissipation, since qubits are designed at the quantum scale, and the thermal energy can represent a source of noise (interference) in their wavefunction. Discrete energy spectra are  typically  very  sensitive  to  small  perturbations  that  can  break  symmetry-related  degeneracies.  Notably, in order to prepare (reset) and retrieve the information of a qubit its quantum state must be destroyed, an operation that necessarily entails some heat release. Until now, such a problematic (a.k.a. ensemble of connected problems) has received a comparatively little attention, because of the already outstanding issues represented by noise from imperfect control signals, interference from the environment and unwanted interactions between qubits, the need for quantum error correction, and the need for operating at cryogenic temperatures. However, questions about temperature, entropy, work, heat, take a very peculiar angle, when seen in the context of quantum mechanics and notably of quantum computing.

\smallskip

The purpose of this article\footnote{An extended and updated version of a series of lectures given between January and April 2022, at the Quantum Information Working Group in the University of Lille, France.}, halfway between a review and a primer, is to give a concise summary of the emerging field of quantum thermodynamics in relation to quantum computing. I should maybe provide a more precise definition at the outset, since it may appear an oxymoron to put together in the same sentence the word "thermodynamics", that is the phenomenological theory of the average macroscopic behavior of heat and work exchanges, and the word "quantum", that in itself represents the epitome of the microscopic world. The two major shortcomings when trying to apply thermodynamics to the quantum domain should be (1) the fact that, by its proper definition, thermodynamics does not contain microscopic information, nor does it have a protocol to relate to the microscopic degrees of freedom; and (2) the fact that it describes only equilibrium states. The first one can be circumvented by passing to the statistical mechanics formulation of thermodynamics, which provides the proper equations and language to make the link with the microscopic. The solution to the second one can leverage on the developments of stochastic thermodynamics, which uses stochastic variables (thus offering a link with the quantum-mechanical notion of probability) to describe the non-equilibrium dynamics typically observed at the molecular length and time scales.

Quite obviously, quantum thermodynamics covers more general questions than just quantum information. Machines that convert heat into electrical power at a microscopic level, where quantum mechanics plays a crucial role, such as thermoelectric and photovoltaic devices, are well known examples of systems requiring the new language of quantum thermodynamics. It is often said that such machines differ from conventional machines by having no moving parts; however, while they may have no \textit{macroscopic} moving parts, they function with steady-state currents of microscopic particles (electrons, photons, phonons, etc.) which are all quantum in nature. Nanotechnology has significantly advanced efforts in this direction, offering unprecedented control of individual quantum particles. The questions of how this control can be used for new forms of heat-to-work conversion has started to be addressed in recent years.\cite{casati17,dutta21}

Hence, our title starts from quantum computers and moves to quantum computing, stressing the fact that to realize a quantum computation you first need to build a physical quantum machine. (Not so obvious, because one can also try to simulate the quantum computation on a classic computer.) The role of thermodynamics, and notably of entropy, therefore will play a dual role in this context, in that it affects both the physical system \textit{and} the computation that is being carried out on that system. Entropy will have a special position, since its different definitions seem to start from rather different premises each time, but eventually end up to very similar, if not formally identical, formulations. We will ask whether a formal similarity also implies, and to what extent, physical identification between different definitions.

\smallskip

This contribution is organized as follows: Sections 2 and 3 give a rapid overview (necessarily incomplete and partial, given the obvious limitations of space) of quantum computers and quantum computing, just some basic details to make this article self-contained for the materials that follow; I will mainly focus as a representative example on superconducting loops and the TransMon that, at this stage of development, seemingly represent the most popular choice of constructors; Section 4 recalls some notions of basic thermodynamics in the context of digital computers; Section 5 deals with the link between thermodynamical entropy and information entropy, and the (ir)reversibility of quantum computations; Section 6 is both the longest and the richest of possibly novel ideas, discussing the reformulation of some classical thermodynamics concepts in the framework of quantum mechanics, and providing several examples of key questions in which quantum thermodynamics can make an impact on quantum computing. Some conclusions and outlook are given in the final Section 7.

\section{On the advantage of quantum computing}\label{sec2}

Quantum advantage (the old word to indicate the computational gain of a quantum device compared to a classical digital computer was "supremacy", but it is currently replaced by "advantage", or "computational advantage") typically refers to situations in which information processing devices built on the principles of quantum physics, attempt at solving computational problems that are not tractable by classical computers.
The resulting quantum advantage is usually defined as the ratio of classical resources required to solve the problem, such as time or memory, to the associated quantum resources. Notably, the numerator in this ratio is often just an estimation, since the problems that are faced are by definition beyond the reach, or at the limits of the capabilities of classic digital computers. Quantum thermodynamics will offer an interesting additional perspective, by comparing also energy dissipation between classical and quantum computing.

Maybe some points need to be clarified, to start with. A quantum computer can use bits and logic gates, just as a digital computer does, therefore at least in theory it should be able to do any computation a classical computer can do, plus a number of other computations that are beyond classical. From the standpoint of computability theory, the key difference between the two is in the state of the bits at any stage of the computation: a classical bit is always in either one of two defined states; a quantum bit is always in a combined state overlapping with the states of (some or all) the other bits. In this way, the interference among quantum bits creates stronger correlations than allowed by classical probability rules,\cite{rau09,quantlog} and can force some bit combinations to be more likely than others.

Measuring a quantum state implemented on a quantum computer, however, will return only classical information, that is, strings of binary code. Then, how can we be sure that a quantum computation was carried out, and not a classical one? Well, the first and simplest check would be to execute many times the same computation. Since quantum computers operate on probabilistic principles, the answer should not be unique but rather a distribution of occurrences, with one being (hopefully) the most probable. Any simple quantum operation, such as summing two bits, necessarily gives a probabilistic answer. See for example the original quantum full adder\cite{feynman85} and its optimized versions:\cite{maslov08} even the best implementation gives a fidelity of 83.333\%.\cite{maslov19}  So, in this case the constant and deterministically repeatable answer of the classical computer is definitely preferable. A recent work \cite{tindall} proved that, by a judicious restructuring of the classical algorithm, even a laptop can outperform the noisy results of the quantum computer on a problem for which an exact solution can be calculated as reference, such as the short-time evolution of the 2D-Ising model.

Therefore, the real challenge would be to propose to the quantum computer a problem that is known to be unsolvable for the classical computer. Beware of the fact that here we intend a class of problems, not a particular instance of the class. "Factorization" is a class of problems, "factorizing the number 4321" is an instance: a classical or quantum algorithm could be good at solving a particular instance, but we are better interested in algorithms that solve the entire class. 

In computation theory, problems whose solution can the obtained in a time that is some power of the size (that is, resources, number of bits, energy) are called polynomial, or P. Given enough resources, a classical computer like a Turing machine can solve any of these P-problems. By contrast, problems that in the general case cannot be solved in a time that grows at best polynomially with the size, are called NP (yes, we are dividing the world into elephants and non-elephants). Factorization of integer numbers is the most typical problem of this kind. A classical computer cannot decompose into prime numbers an integer of arbitrary size, since it would run out of resources at a rate faster than the growth of the required integer. (The largest number factorized, RSA-250 with 795 bits, took the equivalent of 2,700 years on a big supercomputer.\cite{boudot}) 

Already 30 years ago, Peter Shor proposed a quantum-mechanical algorithm that reduces to P-class the NP complexity of factorization, if implemented on an ideal quantum computer.\cite{shor1,shor2} Since then, his algorithm has been programmed on a few different quantum computers, e.g. to factor the number 15 already several times, and more recently the number 21 by using an iterative algorithm to limit the number of necessary qubits.\cite{ventuno} In fact, Shor's algorithm to factor an odd integer $N$ requires a work register with $\log_2$$N$ qubits, plus an output register with $m$ qubits for a precision of $m$ digits: the result will appear as a series of probability peaks in the output register, the narrower and higher, the larger is $m$. The total of resources required is not impressive, but quantum computers with more than a few qubits are still difficult to build, one big problem being the growth of the error rate with the number of entangled qubits. The most recent attempt at factorizing the number 35 on the IBM Q-System-One failed just because of error accumulation.\cite{ibmfail}

Next to problems of deterministic complexity (P, NP, NP-hard, NP-complete), a class of bounded-error probabilistic-polynomial problems, or BPP, has been defined. These are problems that can be solved in a polynomial time and include random processes (such as in a probabilistic Turing machine), but are bounded, meaning that the algorithm gives the right answer with a large probability, fixed at 2/3. Obviously, problems that are in the P class are also in the BPP class, and it is believed (but not proven) that the two classes coincide, especially after it was demonstrated that the primality problem (i.e., determining whether a number is prime) is also in P.\cite{agrawal} 

By lowering the requirement of giving the correct answer to more than 1/2 probability (that seems just one inch above the tossing of a coin to find the answer), the larger class of PP (probabilistic-polynomial) problems is defined. And somewhere between the two, there is the BQP class, or bounded-error quantum-polynomial complexity.\cite{bernstein}   By definition. BQP contains all BPP problems, and obviously all P problems; it also includes some NP problems, such as factorization. And it includes some problems "beyond-NP", that is, problems for which a classical computer cannot find, or even check the correctness of, the answer in a polynomial time. An example is the boson sampling problem, in which somebody wants to determine the probability distribution of an ensemble of $M$ identical, non-interacting bosons (photons, spin-0 atoms\footnote{In these experiments, atoms such as $^{87}$Rb or $^{133}$Cs are used, which do not look like bosons. At low temperatures, however, electrons and the nucleus behave as a unique ensemble, so that the Rb  spin-3/2 nucleus plus the 37 spin-1/2 electrons make up a single system with integer spin.}) after scattering through an interferometer.\cite{aaronson} Such a physical experiment requires a mathematical tool to calculate the answer, the \textit{permanent} of an $M\times M$ matrix, which would normally require an exponential time to compute ($\mathcal{O}(M^2 2^M)$, \cite{marvin}) on a classical computer. Therefore, this seems like an ideal case in which to test the advantage or "supremacy" of a quantum computer with respect to classical, digital computers (see below).

\bigskip 

\addcontentsline{toc}{subsection}{What do quantum computers look like, A.D. 2024?}

\noindent \textbf{What do quantum computers look like, A.D. 2024?} The hardware requirements to achieve computational advantage can be summarized by three key properties:
\begin{itemize}
\item the quantum systems must initially be prepared in a well-defined pure state;
\item arbitrary unitary operators must be available and controllable in order to launch an arbitrary entangled state; 
\item measurements (read-out) of the qubits must be performed with high quantum efficiency.
\end{itemize}
We can rank the different solutions proposed up to date in about three broad classes:
 
 \smallskip
 
\textit{Gate-based quantum computation} and the related class of universal digital algorithms, are approaches that rely on a quantum processor, encompassing a set of interconnected qubits, to solve a computation that is not necessarily quantum in nature. The dominant technique for implementing single-qubit operations is via microwave irradiation of a superconducting loop (see below, Sect.3). Circuit quantum electrodynamics (CQED), the study and control of light-matter interaction at the quantum level,\cite{blaisrmp} plays an essential role in all current approaches to gate-based digital quantum information processing with superconducting circuits.
Electromagnetic coupling to the qubit with microwave pulses at the qubit transition frequency drives Rabi oscillations in the qubit state; control of the phase and amplitude of the drive is then used, to implement rotations about an arbitrary axis of the quantum state of each qubit; these are the logical gates, which perform the sequence of required operations in the algorithm as a sequence of unitary transformations of the state of the ensemble of qubits. 
Typically, current universal algorithms are tailored to a specific, and potentially noisy hardware (noisy intermediate-scale quantum-, or NISQ-technology \cite{nisq,eddins}) in order to maximize the overall fidelity of the computation, despite the absence of a yet complete and reliable scheme for quantum error correction. (See for example the reviews in \cite{oliver20, china20}. More on this class of devices in the following.)

\smallskip 

\textit{Adiabatic quantum computation} is an approach formally equivalent to universal quantum computation, in which the solution to computational problems is encoded into the ground state of a time-dependent Hamiltonian. Solving the problem translates into an adiabatic (that is, very slow) quantum evolution towards the global minimum of a total-energy landscape that represents the problem Hamiltonian. Compared to numerical annealing on a classical computer, achieved by using simulated thermal fluctuations to allow the system to escape local minima (such as in the kinetic Monte Carlo method), in quantum annealing the transitions between states are caused by quantum fluctuations, rather than thermal fluctuations, leading to a highly efficient convergence to the ground state for certain problems. The D-WAVE quantum annealers are a successful line of development of this scheme, having already demonstrated the successful operation of a machine with more than 2,000 superconducting flux qubits, on real physics problems (see e.g. \cite{harris18}). More generally, any optimization problems that can be reframed as minimization of a cost function (the "total energy") could be efficiently run on such devices. It may be worth noting that up to now, there is still no firm proof that an adiabatic computer could offer an effective speed-up (advantage) over an equivalent classical computation.\cite{ronnow,raponi}

\smallskip

\textit{Quantum simulators} are well-controllable devices that mimic the dynamics or properties of a complex quantum system that is typically less controllable or accessible. The key idea is to study relevant quantum models by emulating or simulating them with an hardware that itself obeys the laws of quantum mechanics, in order to avoid the exponential scaling of classical computational resources. 
Quantum simulators are problem-specific devices, and do not meet the requirements of a universal quantum computer. This simplification is reflected in the hardware requirements and may allow for a computational speed-up with few, even noisy quantum elements, for example by emulating specific Hamiltonians and studying their ground state properties, quantum phase transitions, or time dynamics (see e.g. \cite{fitz17,ma19}). Therefore, quantum simulators might be ready to address meaningful computational problems, demonstrating quantum advantage well before universal quantum computation could be a reality.
 
 \bigskip
 
 \addcontentsline{toc}{subsection}{Chariots of fire}

\noindent \textbf{Chariots of fire.}
The first claim of quantum computational advantage was launched in 2019, with the general-purpose Sycamore quantum processor, housing 54 superconducting programmable TransMon qubits operating at 10 mKelvin, built by a team of Google engineers in Santa Barbara, CA.\cite{sycam} 
The qubits are arranged in a rectangular $9\times 6$ Ising layout, with gates operating either on single qubits, or pairs of neighbor qubits. Pseudo-random quantum circuits are realized by alternating single-qubit and two-qubit gates, in specific, semi-random patterns. This gives a random unitary transformation perfectly compatible with the hardware.
The circuit output is measured many times, producing a set of sampled bit-strings. The more qubits there are, and the deeper the circuit is, the more difficult it becomes to simulate and sample these bit-string distributions on a classical computer. By extrapolation, Google engineers estimated that a sampling that required about 200 seconds run
on Sycamore would have taken $\sim$10,000 years on a million-core supercomputer.
However, the claim was questioned by a team at IBM (whose quantum computer Q-system One is also based on superconducting qubits) who devised a much faster classical algorithm, based on which they predicted (yet without performing it) that the classical calculation would get down to about 2-3 days. Two years later, a Chinese team used a tensor-based simulation on their Sunway digital supercomputer, to perform the same simulation in 304 seconds.\cite{sunway} 

\bigskip

 \addcontentsline{toc}{subsection}{The Dragon Labyrinth}

\noindent \textbf{The Dragon Labyrinth.}
It has been demonstrated already some time ago that by using only linear optical elements (mirrors, beam splitters, phase shifters) any arbitrary 1-qubit unitary operation, or equivalent quantum gate, can be reproduced.\cite{knill2001} 
The flip side of the coin is that photon-based implementation is not a very compact architecture. However, photonic quantum microcircuits are under active development (see the Canadian startup \textit{Xanadu}\cite{xanadu}).
The boson random sampling experiment, proposed by Aaronson and Arkhipov at MIT,\cite{aaronson} entails calculating the probability distribution of bosons whose quantum waves interfere by randomizing their positions. The probability of detecting a photon at a given position (e.g. behind a diffraction grating) involves practically intractably big matrices (the “Torontonian” requires to compute $2N$ determinants of rank $N$)
The Heifei University group published results of their largest optical quantum computer.\cite{pan2021} They used 72 indistinguishable single-mode Gaussian squeezed states as input, injected into a 144-mode ultralow-loss interferometer, generating entangled photon states in a Hilbert space of dimension $2^{144} \simeq 10^{43}$.
The classical solution requires calculating $10^{43}$ determinants $144 \times 144$, so the “supremacy” in this case is clearly undisputable.

\bigskip

\section{Superconducting qubits and quantum gates }\label{sec3}

A simple search on the internet will present you a fairly large number of options to realize a quantum bit, or \textit{qubit}, ranging from cold atoms,\cite{winter23} to trapped ions,\cite{bruze19} to nuclear spins, quantum dots, topological systems with a gap, and others.  However, when it comes to practical implementations on currently existing quantum computers, the long list turns into a rather short one. Basically, only three choices along with some variants, are the ones found along the beaten path: superconducting quantum dots, ion traps, and photonic circuits. To avoid excessive length, this article will not deal with photon-based computing. Aside of the long tradition that makes generation, control and measurement of photons as quantum systems a routine in many laboratories and industry, and despite their many advantages, among which working at room temperature, photon-based quantum computing has one major disadvantage, in that photons do not interact with each other. This key issue requires a special, dedicated approach to the problem, which is well described in a number of excellent reviews already (see e.g. \cite{pryde19,roadmap,sciarrino}).  Hence, since this is not primarily a review on quantum computing hardware, I will briefly discuss in this Section as a practical example of physical implementation only the superconducting (SC) qubits, which are up to now the most popular choice (IBM, Google, Rigetti, and others), in its two main variants, the charge and the flux qubit. 

\bigskip

 \addcontentsline{toc}{subsection}{TransMon basics}

\noindent \textbf{TransMon basics.} The basic idea behind the SC qubit is to create a tunable oscillator in the solid-state with well defined quantum mechanical states, between which the system can be excited by means of an external driving force. The quantum harmonic oscillator is a resonant circuit that can be schematized as a typical LC-circuit-equivalent, with characteristic inductance $L\simeq$1 nH and capacitance $C\simeq$10 pF, which result in a resonant frequency:

\begin{equation}
f = \frac{1}{2 \pi \sqrt{LC}} \simeq 1.6 \text{\,\,GHz, and\,\,\,\,\,} \lambda = \frac{c}{f} \simeq 2 \,\,\text{cm}
\bigskip
\end{equation}

In order to address quantum states individually a non-linear component to the circuit  must be introduced, thus making it an anharmonic circuit. A Josephson junction (J-J) is a device that consists of an insulator “sandwiched” between two superconductors, and can act as a non-dissipative and non-linear inductance. For this purpose, temperature must be in the mK range, sufficiently low for electrons to condense below the Fermi energy and form Cooper pairs.

Since the dimensions of the J-J are of only a few hundred $\mu$m (i.e., much less than the circuit’s operating wavelength $\lambda$ above), everything falls well within the lumped-element limit, and it can be described by using one collective degree of freedom $\Phi$ (the magnetic flux). In its basic implementation, the SC qubit can be designed in different manners: (1) as a \textit{charge} qubit, composed of a J-J and a capacitor;  adjusting the voltage can control the number of Cooper pairs; (2) as a \textit{flux} qubit, with a loop inductance replacing the capacitor; changing the bias flux can adjust the energy level structure;  (3) as a \textit{phase} qubit, with just the J-J and a current modulator; adjusting the bias current can tilt the potential energy surface.

\begin{figure}[h]
\centering
\includegraphics[width=0.9\textwidth]{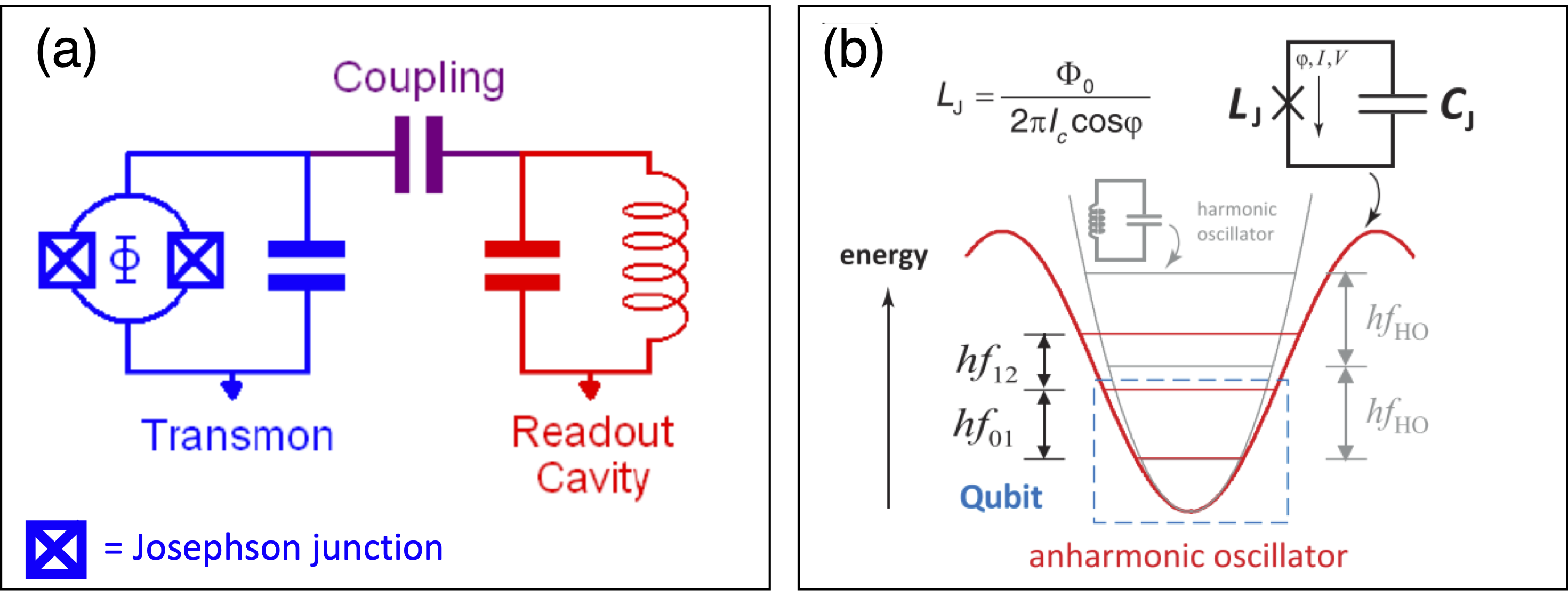}
\caption{Basic scheme of a Transmon. \textbf{(a)} The double J-J forming a loop, with its shunting capacity (blue), the read-out $LC$ resonator (it could be replaced by a resonant cavity), and the coupling capacity (grey). \textbf{(b)}  Difference between the linearly-spaced levels of the harmonic oscillator, and the non-linearly-spaced levels in the magnetically driven cosine potential. }
\label{fig1}
\end{figure}

For $\Phi$  to be treated as a quantum-mechanical variable, the width of the energy levels of the resonator must be smaller than their separation, which puts a constraint on the damping $Q$ of the oscillator. Hence, to keep a $Q\gg1$, the inductor could only be made by a superconducting wire: a quantum-LC resonator. But a single quantum-LC is still a harmonic oscillator, which means that its equally-spaced energy levels are not individually addressable. It is then impossible to restrict the system to only two states, as a qubit requires.

However, the magnetic energy in the J-J is not classically quadratic in the flux, but rather proportional to the cosine of $\Phi$, as $E_J=E_J^{max}|\cos(\pi \Phi/\Phi_0)| = L_J I_c^2$. Here $I_c=2E_J/\hbar$ is the critical current (maximum current that can flow coherently through the junction), $\Phi_0 = \hbar /2e$  is the SC quantum, and  $L_J = \Phi_0 / 2 \pi I$  is the J-J inductance. The junction current is $I=I_c \sin\phi$, with $\phi=(2\pi/\Phi_0)\Phi$ the Josephson phase.\cite{blaisrmp,oliver20} 
Due to this anharmonicity (Fig.\ref{fig1}b), the ground and first-excited states of the Cooper pair may be uniquely addressed at a frequency $f_{01}$, typically in the microwave range 4-8 GHz, without significantly perturbing higher-excited states of this “artificial atom”. Then, the two lowest-energy states make up an effective two-level system, i.e. a “pseudo-spin-1/2” system, although the SC loop \textit{per se} would not be a true 2-level system.

\bigskip

 \addcontentsline{toc}{subsection}{Charge qubit}

\noindent \textbf{Charge qubit.} Starting from the three basic implementations, different and more advantageous types of SC qubits have been invented, the TransMon being by far the most popular. Based on the charge-qubit model, in which the number of Cooper pairs $N$ is the main variable, a flux-tunable transmon can be realized with double J-J (a modified SQUID loop, first proposed by J. Koch's group in 2007,\cite{Koch2007} Fig.\ref{fig1}a) that can store an energy $E_J = (\Phi_0/2\pi)I$. 
The circuit is shunted by a large coupling capacitance, such that the coupling energy $E_C = e^2/2C \ll E_J$, thus giving a large $Q$.
The advantage of such a double-J-J is that the values of magnetic flux $\Phi$ can be fine-tuned, and each qubit can be individually addressed by a ”gate” (actually, a sequence of GHz pulses).
The state of the qubit is readout by a second resonator (“cavity”) whose resonant frequency $f_R$ is chosen to be far from $f_{01}$. 
Then, the two possible qubit states show up as a (small) red- or blue-shift $\Delta$ about the central readout frequency, $f_R \pm \Delta$.

The Hamiltonian of the classical equivalent circuit can be written as:

\begin{equation}
\mathcal{H} = 4 E_C \left(  N - \tfrac{1}{2} \right)^2 - E_J \cos \phi + \underline{ W(t) \sin (\Omega_R t + \phi) }
\bigskip
\end{equation}

The quantum version for the TransMon qubit (such as found in IBM Quantum System, or Google Sycamore) runs on a slightly modified version of the celebrated Jaynes-Cummings theoretical model,\cite{lo98} and looks like:

\begin{equation}
\mathcal{\hat{H}} = \hbar \sqrt{8 E_C E_J} \left( a^{\dagger}a  + \tfrac{1}{2} \right) - \hbar E_C \left( a{\hat{\sigma}}^+ + a^{\dagger}{\hat{\sigma}}^- \right) + \
\underline{ \Omega_R \left[ \mathcal{J}(t){\hat{\sigma}}_x + \mathcal{Q}(t){\hat{\sigma}}_y \right]  }
\bigskip
\end{equation}

In the two equations, the underscored parts represent the external field (a train of microwave pulses) driving the Hamiltonian.
$N$ is the number of excess Cooper pairs (destroyed/created by the operators $a,a^{\dagger}=\sqrt{n}\ket{n\mp 1}\bra{n\pm 1}$; $\hat{\sigma}_{x,y,z}$ and $\hat{\sigma}^{\pm}=\hat{\sigma}_x\pm i\hat{\sigma}_y$ are Pauli matrices; $\Omega_R$  is the Rabi frequency; $\mathcal{J}$ and $\mathcal{Q}$ are the in-phase and quadrature components of the microwave signal $W(t)$. In principle, $N$ and $\phi$ are both good quantum numbers to describe the TransMon; however, in the $E_C \ll E_J$ limit only $N$ is well defined, while $\phi$ fluctuates randomly.

\bigskip

 \addcontentsline{toc}{subsection}{Flux qubit}

\noindent \textbf{Flux qubit.} A variant of the transmon is realized with a SC ring interrupted by 3 or 4 Josephson junctions.
The qubit is engineered so that a persistent current flows continuously when an external magnetic flux is applied. Only an integer number of flux quanta penetrate the SC ring, resulting in clockwise or counter-clockwise mesoscopic supercurrents (typically 300 nA) in the loop, which compensate (screen or enhance) a non-integer external flux bias. The $\phi$ degree of freedom becomes now the main variable, the number of flux quanta $N$ being random, and the coupling energy dominates over the charging energy, $E_J \ll E_C$.

When the applied flux through the loop is close to a half-integer number of flux quanta, the two lowest-energy loop eigenstates are found in a quantum superposition of the two currents.
This is what makes the flux qubit a spin-1/2 system, moreover with separately tunable $z$ and $x$ fields, 

The flux qubit has been used as building block for quantum annealing applications based on the transverse Ising Hamiltonian.\cite{hauke2020}     
A  typical quantum Hamiltonian that can be implemented in a connected network of flux qubits, such as in the D-WAVE Chimera or Pegasus architectures, looks like:

\begin{equation}
\mathcal{\hat{H}} = \Lambda(t) \left[ \sum_i h_i \hat{\sigma}^z_i + \sum_{i<j} J_{ij} ( \hat{\sigma}^z_i \cdot \hat{\sigma}^z_j ) \right] + \Gamma(t) \sum_i \Delta_i \hat{\sigma}^x_i
\bigskip
\end{equation}

The $h_i$ are asymmetry energies, $J_{ij}$ represent the coupling matrix elements, and $\Delta_i$ are tunneling energies. At the beginning of the quantum annealing process, it is $\Gamma(0)$=1 and $\Lambda(0)$=0, to create a known ground state as an equal superposition in the computational basis.  During the adiabatic annealing protocol, the two parameters slowly evolve towards $\Gamma \rightarrow 0$ and $\Lambda \rightarrow 1$. 

\smallskip 

Other transmon variants have been proposed to counter some of the practical problems encountered in the different SC loops implementations, such as the C-shunt flux qubit,\cite{you07} to reduce charge noise; the "fluxonium",\cite{manu09} to address the noise from inductance and offset charge; the "0-$\pi$" qubit,\cite{kitaev,gyen21} designed to improve the symmetry of the two current states; and various types of hybrid qubits (see, e.g., \cite{marcos10,kubo10,zhu11}), in which the SC loops are coupled to solid-state elements, a doped crystal, or a resonant cavity, to exploit the advantages from different quantum effects. Over the recent years, the key objective of increasing the lifetime of the qubit state has been pursued, extending the coherence time from mere fractions of $\mu$s well into the ms domain.\cite{pop14}

\smallskip 

The next important operation to consider is the read-out of the information from the qubit. For the SC loops, it is possible to detect either charge, flux, or inductance. The most popular method is the dispersive read-out,\cite{wall04} in which the qubit and the resonator (see again Fig.\ref{fig1}a) are coupled by a strength parameter $g\ll\Delta=\omega_{01}-\omega_R$, as in the approximate Hamiltonian:

\begin{equation}
\mathcal{\hat{H}} = -\frac{\omega_{01}}{2} \hat{\sigma}^z + \left( \omega_R + \frac{g^2}{\Delta} \hat{\sigma}^z \right) a^{\dagger}a
\bigskip
\end{equation}

The presence of a $\ket{0}$ or $\ket{1}$ state in the qubit shows up as a small frequency shift in the resonator by the quantity $g^2/\Delta$. The read-out is "dispersive" in the sense that the signal corresponding to the two possible states appears clustered in two disjointed clouds in the complex plane.\cite{blaisrmp}

\bigskip

 \addcontentsline{toc}{subsection}{Qubits and quantum gates}

\begin{figure}[h]
\centering
\includegraphics[width=0.8\textwidth]{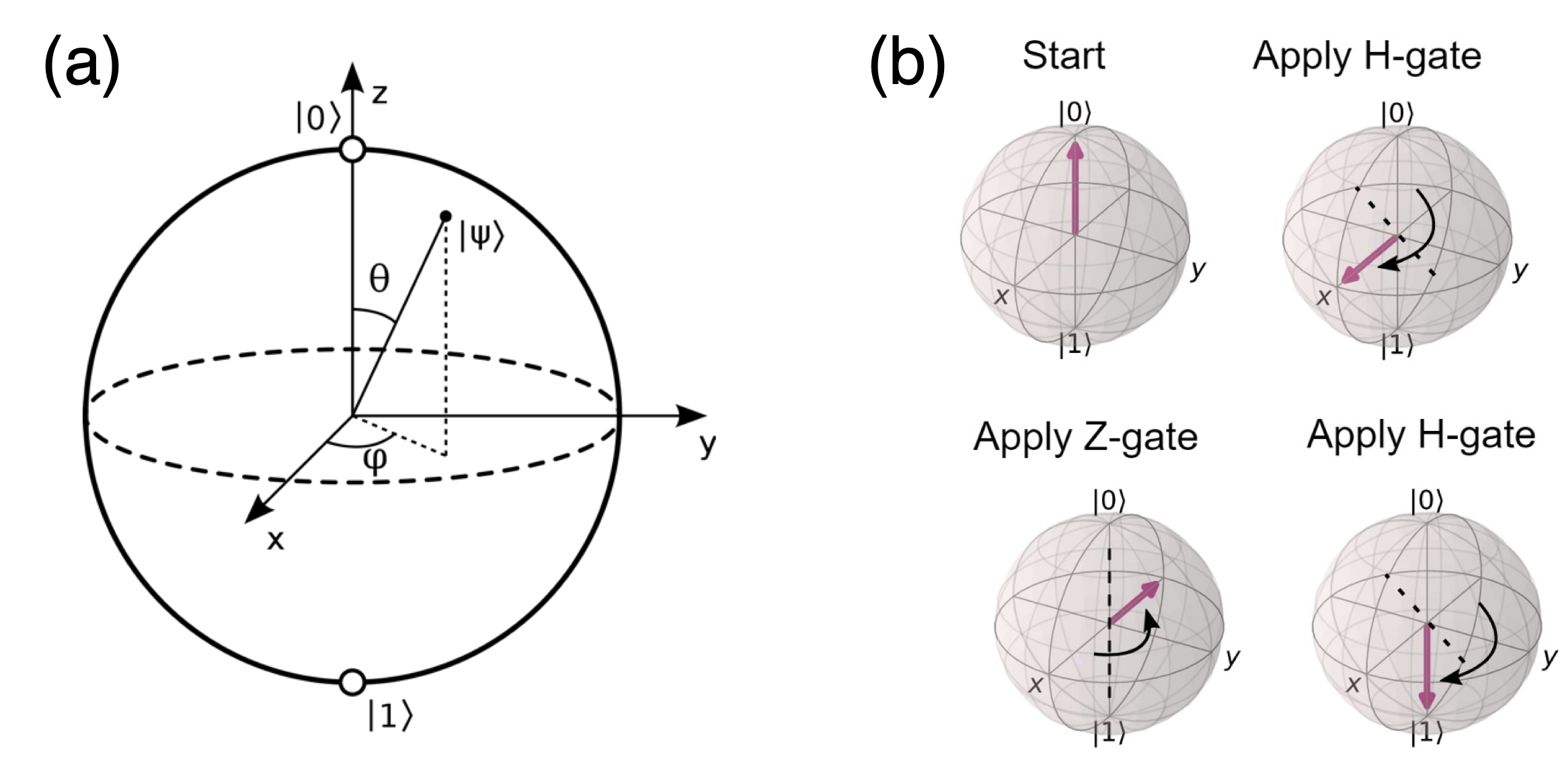}
\caption{ \textbf{(a)} The Bloch sphere. \textbf{(b)} Schematic of the transformation on the Bloch surface corresponding to the sequential application of a $H$, $Z$ and again a $H$ gate, resulting in the spin flip from $\ket{0}$ to $\ket{1}$ (therefore equivalent to an $X$ gate). }
\label{figbloch}
\end{figure}

\noindent \textbf{Qubits and quantum gates.}
Independently on their actual physical implementation, qubits are mathematically defined as two-state quantum systems, described by a state vector in a 2-dimensional Hilbert space, spanning a closed surface with conserved
norm (i.e., a sphere, called the "Bloch sphere", Figure \ref{figbloch}a). A standard basis is defined by the two vectors $\ket{0}$ and $\ket{1}$, conventionally aligned with the positive and negative direction of the z-axis.

A quantum computer executes a sequence of unitary transformations, $U_1 … U_n$, as specified by a quantum algorithm, with each transformation acting on one or two (rarely three) qubits. The unitary transformation is executed by a "gate", actually a physical operation that makes an external system (microwave or laser pulse, magnetic field switch, AC/DC signal) to interact with the qubit to modify its state. For a single qubit it is often a rather simple operation, for example a microwave pulse at the exact frequency of the J-J junction. For two-qubit gates it may be more complicated, since qubits are in principle arranged in such a way that do not interact with each other, in their ground state. Coupling can be realized for example by properly designing the capacitance between the pair,\cite{yama2003}  or through their mutual inductance,\cite{hime2006} or by a microwave cavity in which confined photons transfer the quantum state between qubits,\cite{mayer2007} or yet by other means.

One-qubit gates act on either the phase or the excitation energy of a single qubit by applying a rotation on the Bloch surface (Fig.\ref{figbloch}b). The states of the qubit can be represented as column vectors, 
$\ket{0}=\left(\begin{smallmatrix}1 \\ 0\end{smallmatrix}\right)$ and $\ket{1}=\left(\begin{smallmatrix}0 \\ 1\end{smallmatrix}\right)$. 
 Then the simplest operations of transformation of the state (a "gate") can be schematized by 2$\times$2 matrices. An arbitrary rotation of the qubit state about an axis $\hat{\textbf{n}}$ is represented as $U_{\hat{\bf{n}}}(\phi)=\exp(-i\phi\hat{\bf{n}}\cdot\vec{\sigma})=\cos(\phi)\textbf{I}-i \sin(\phi)(\hat{\bf{n}}\cdot \vec{\sigma})$, with $\vec{\sigma}$ the block vector having the Pauli matrices as its \textit{xyz} components.
 
 \smallskip
 
 In the case of the TransMon, standard rotation gates are available (hardware implemented) in the \textit{xy}-plane or in the \textit{z}-axis. For the \textit{xy} single-qubit gate, the Hamiltonian reads:
 
 \begin{equation} 
 \mathcal{H} = - \frac{\hbar}{2} \omega_{01} \sigma_z + W \cos (\omega_R t - \phi) \sigma_x = \\
  - \frac{\hbar}{2} \left[ \Delta \sigma_z - W (\cos\phi\sigma_x + i \sin\phi\sigma_y) \right]
  \bigskip
\end{equation}

When the MW frequency is exactly tuned to the qubit frequency it is $\Delta$=0, and the rotation in the \textit{xy} plane is fixed only by the choice of the phase angle $\phi$. This makes the $xy$ or phase-gate one of the most important elements of quantum computing.

Students may find it difficult to grasp the meaning of the phase gate, since in introductory classes of quantum mechanics the role of the phase remains rather obscure, and is often swept under the carpet by noticing that "phase disappears when taking the $|\psi|^2$ of the wavefunction".\footnote{This may be true for pure states and for many expectation values, such as energy, but it is not always the case. Consider, e.g., the mixed state $\psi=a_1\phi_1+a_2\phi_2$: its $|\psi|^2=|a_1|^2|\phi_1|^2+|a_2|^2|\phi_2|^2 + (a_1^*a_2\phi_1^*\phi_2+c.c.)$ depends on the phase, since $a_1^*a_2=|a_1||a_2|e^{i(\theta_2-\theta_1)}$. The energy of such a mixed state has periodic oscillations at the frequency $\omega=(E_2-E_1)/\hbar$ because of the rotating phase. }
If the output of a measurement is a random distribution, in a classical setting one can only accumulate probability amplitudes of each instance; however, in the quantum calculation, constructive or destructive interference allows to amplify or suppress some of the outputs, by manipulating the phases of the qubits. A great example is the quantum Fourier transform (QFT) which finds periodic instances in a sequence, just like its classical counterpart.  The QFT algorithm transforms a $m$-bit state $\ket{\psi}=\sum_i \alpha_i \ket{i}$ to $\ket{\psi'}=\sum_i \beta_i \ket{i}=\sum_i \sum_k \alpha_i e^{2\pi i \phi_k} \ket{i}$, $i,k=1...m$. It requires the application of one Hadamard gate to each pair of qubits, followed by a sequence of phase-gates for each term in the $k$-sum, resulting in a total of $m$ Hadamard + $m(m+1)/2$ phase-gates (that is, $\mathcal{O}(m^2)$ polynomial complexity).
 

\vspace{6pt}

Two-qubit gates carry out controlled transformations of the second qubit state (\textit{target}), conditioned by the state of the first one (\textit{control}). 
Compared to single-qubit gates, whose working speed is limited by the strength of the driving fields,\cite{frey}  two-qubit  gates (notably, the entangling gate) can only operate at a speed proportional to the interaction strength between the qubits.\cite{ballance,imreh} This is typically weaker than available single-qubit drive strengths and cannot be easily increased, thereby representing one important limit to the coherence time (see below). For the SC qubits, moreover, the limited anharmonicity makes single-qubit gates not much faster than two-qubit gates.\cite{zajac21,moska22}

As the two-qubit state is a column vector of dimension 4, the gates can be written as a 4x4 matrix. The controlled transformations can be of two main types:
the Controlled-NOT (CNOT), which leaves $\ket{00}$ and $\ket{01}$ unaltered, and swaps $\ket{10}$ and $\ket{11}$;
the Controlled-Phase (CPHASE), which flips the phase of the two-qubit state if both the qubits are excited. It is interesting to note that CNOT  can be constructed by applying Hadamard to the target qubit, then CPHASE to this new state, and Hadamard again (see e.g. \cite{lantz05}).
 The ensemble of CNOT gate, the Hadamard, and all phase gates, form an infinite set of \textit{universal} gates, by which any $m$-qubit unitary operation can be represented using $\mathcal{O}(m4^m)$ such gates.

\bigskip
 \addcontentsline{toc}{subsection}{Entanglement}

\noindent \textbf{Entanglement.} It is worth noting once more that \textit{coupling} is not the same thing as \textit{entanglement}. The coupling refers to the physical mechanism allowing the exchange of information between different qubits. The result can be a non-entangled, a partial, or a maximally entanglement of the state vector $\ket{\psi}$,\cite{franco2016} which can be detected by looking at the quantity $\rho=\ket{\psi}\bra{\psi}$: if $\rho^2=\rho$ then the state is pure and entangled. 
For example, a two-qubit state like
$\ket{\psi}= \tfrac{1}{\sqrt{2}} \left( \ket{00} + \ket{01} \right)$
is separable (non-entangled), because it can be written as 
$\ket{0} \otimes \tfrac{1}{\sqrt{2}} \left( \ket{0} + \ket{1} \right)$; therefore, it is $\rho=\tfrac{1}{2}(\ket{00}+\bra{01})^2$,
but $\rho^2=\sum_{i,j} \ket{i}\braket{i|j}\bra{j}=$ $\tfrac{1}{2}(\ket{00}\bra{00}+\ket{01}\bra{01})$, because of the orthonormality of the $\ket{ij}$ set.\cite{caban}

\smallskip

Maximally-entangled states are called "Bell states". The meaning of maximum entanglement is usually taken as the maximization of Von Neumann's entropy (see below, Eq.(\ref{eqvonn}) and on). A more layman interpretation is that the state is described by a single wavefunction (i.e., not separable), so that a measurement of any qubit gives the values of all the others deterministically. (By contrast, a mixed state would give a statistical mixture of all qubits.) 

Entangled states can be obtained by the sequential application of a Hadamard single-qubit gate, followed by a CNOT. The Hadamard produces a 50/50 superposition of the basis states, for example:

\begin{equation}
[H]  \ket{0} = \frac{1}{\sqrt{2}} \left( \begin{matrix} 1 & 1 \\ 1 & -1 \end{matrix} \right) 
\left( \begin{matrix} 1 \\ 0 \end{matrix} \right) = \frac{1}{\sqrt{2}} \left( \begin{matrix} 1 \\ 1 \end{matrix} \right) =
\frac{\ket{0}+\ket{1}}{\sqrt{2}}
 \bigskip
\end{equation}

 Then, the CNOT gate operates on the product states, such as $\ket{\psi_p^{\pm}}=\tfrac{1}{\sqrt{2}}(\ket{0} \pm \ket{1})\otimes\ket{p}$, $p=0,1$ to obtain the finally entangled states. For example:

\begin{equation}
[CNOT] \ket{\psi_0^+} =\frac{1}{\sqrt{2}} \left( \begin{matrix}  1 & 0 & 0 & 0 \\
                                                          0 & 1 & 0 & 0 \\
                                                          0 & 0 & 0 & 1 \\
                                                          0 & 0 & 1 & 0  \end{matrix} \right)
                                    \left( \begin{matrix}  1 \\ 0 \\ 1 \\ 0 \end{matrix} \right) =
 \frac{1}{\sqrt{2}}  \left( \begin{matrix}  1 \\ 0 \\ 0 \\ 1 \end{matrix} \right) =
  \frac{1}{\sqrt{2}}  \left( \ket{00} + \ket{11} \right) 
\bigskip
\end{equation}

For two qubits, the four possible Bell states obtained by combining all product states are given by:

\begin{equation}\begin{split}
&\ket{\psi_B^{\pm}} = \tfrac{1}{\sqrt{2}} \left( \ket{00} \pm \ket{11} \right) \\
\\
&\ket{\phi_B^{\pm}} = \tfrac{1}{\sqrt{2}} \left( \ket{01} \pm \ket{10} \right) \\
\end{split}\end{equation}

However, not all entangled states are Bell states. For example, the states $\ket{\psi}=\cos(\theta) \ket{00} + \sin(\theta) \ket{11}$ for
$\theta \in (0,\pi/4)$
 are all entangled but are not Bell-states.

\smallskip

A last point worth noting, is that the time profile of the interaction Hamiltonian is controlled by classical parameters, such as the intensity of a laser beam, the value of the gate voltage, or the current intensity in a wire. Of course, all such parameters are also quantum-mechanical in nature, when examined at the atomic level; the fact that they behave classically means that there should be no entanglement between their (very large) quantum states, and the internal states of the qubits of the quantum computer.

\bigskip

 \addcontentsline{toc}{subsection}{Decoherence, dephasing, thermalization}

\noindent \textbf{Decoherence, dephasing, thermalization.}
However, all physical quantum systems are subject to decoherence and dissipation, mainly arising from their noisy interaction with the environment. As we will see later (Section 5), when exploring the connection between thermodynamics and information, any realistic sequence of operations of a quantum information processing device is irreversibly accompanied by the production of  entropy, which pairs with the irretrievable loss of (quantum) information into the environment.  Then, some questions immediately appear: 
\begin{itemize}

\item What are the physical limitations on information processing set by thermodynamics?

\item Can we maintain quantum computers in the deep quantum regime, so that we can actually exploit their advantage w/r to classical computers?

\item The exponential increase in computing capability will entail and exponential increase of thermodynamic work and dissipated heat?
\end{itemize}

Within a standard picture for spin-1/2 systems, there are two characteristic decay rates that contribute to coherence loss: $\Gamma_1=1 / T_1$ is the  longitudinal relaxation rate (an energy decay rate), that is the time over which the qubit exchanges energy with its environment; $\Gamma_2 = 1/T_2 = 1/(2T_1 + 1/T_{\phi})$  is the transverse relaxation rate (a decoherence rate), that is the time over which the device remains phase-coherent. 
In simpler words, $T_1$ is the time taken by a qubits to decay spontaneously e.g. from $\ket{1}$ to $\ket{0}$, while $T_2$ is where a qubit dephases into a mixture of states such that the phase can no longer be accurately predicted.
Over the past 20 years, a steady increase in $T_2$  has brought superconducting qubits from the stage of laboratory experiments, to the capability of building the first quantum computers.\cite{olivermrs,gil2020} 

Currently the error rate on the best quantum computers is about 1\% for each elementary operation. Although a 99\% accuracy  may seem already high, a single mistake affects the whole entangled system: just one error corrupts the result. One way to improve errors could be to replicate $N$  identical copies of the logical unit and have them "vote" on the output. Only if all $N$ physical qubits give the same answer, the logical qubit is correct. (This is similar to what happens in classical computers, e.g. with the Hamming correction code.) 

Another method that is becoming standard is the introduction of correction, or "ancilla", qubits, with the same logic of the parity bit in classic digital computers.
The ancilla qubit is prepared in $\ket{0}$, and then a sequence of CNOT gates are applied, from the working qubits onto the ancilla qubit. These gates flip the ancilla or "check bit" between $\ket{0}$ and $\ket{1}$ an even or odd number of times, depending on the parity of the bit string stored in the data qubits. When the ancilla qubit is measured, the parity of the state is the only thing that is measured, without interfering with the rest of the quantum computation.

However, as the number of logical qubits grows, the number of layers to correct the original plus the correction qubits grows exponentially.
Google's labs estimate is that current technology may require 1,000 physical qubits to encode 1 logical qubit and attain an error rate of 1 in $10^9$.  

Introduced as a measure of the practical estimate of the minimal availability of quantum resources to perform a computation, the \textit{quantum volume} of a quantum computer depends on the number of qubits $N$, as well as the number of steps that can be executed while remaining in a coherent state, that is the \textit{circuit depth}, $d$:

\begin{equation}
V_Q = \min[N,d(N)] 
\bigskip
\end{equation}

The variation of $d$ with the number of qubits is $d(N)\simeq 1/(N\epsilon)$, for an average error rate $\epsilon$. However, a quantum algorithm typically engages subsets of $n$ qubits from those available in the whole machine $N$. IBM's modified definition of quantum volume \cite{moll18} is the equivalent complexity to simulate the same quantum circuit on a classical computer:

\begin{equation}
\log_2 V_Q = \arg \underset{n \leq N}{\max} \{ \min [n,d(n)] \}
\bigskip
\end{equation}

Such definitions only provide a measure of the theoretical feasibility of a computation, neglecting other constraint factors as, e.g., read-out times, 1/$f$ noise, quantum error correction, magnetic or current (phase) fluctuations.

\section{Thermodynamics in a classic digital computer}

Thermodynamics was developed in the XIXth century, to provide a unified framework between mechanical sciences and thermometry. 
At the time, the motivation was very practical, namely use temperature to put bodies into motion - as clearly indicated by its name. 
In other words, the goal was to design and optimize thermal engines, i.e. devices that exploit the transformations of some “working substance" to convert heat into work. 
Work and heat are two ways to exchange energy, according to the First Law of thermodynamics it is possible to convert one into another. 

However, turning heat into work and back into heat,  comes at a cost: it is not possible to cyclically extract work  from a single hot bath (Kelvin 1851)
This no-go statement is one of the expressions of the Second Law of thermodynamics, which ultimately deals with irreversibility. 
The concept of work came in origin from mechanical sciences (Lazare Carnot, 1803) and represents a form of energy that can be exchanged reversibly: in principle, there is no time arrow associated with work exchanges (at least for conservative forces), since the equations of motion in classical mechanics are perfectly time-reversible.
But when building steam machines it is always found that heat $Q$ spontaneously flows (only) from hot to cold bodies. To extract work $W$, a source and a sink at different temperatures $T_1$ and $T_2$ are necessary, independently on the nature of the exchange medium, as stated by Sadi Carnot in 1823 in a idealized experiment, the "Carnot cycle", for which he derived a theorem regarding the efficiency of a machine producing work.

A Carnot cycle is a closed ensemble of operations by which a thermodynamic machine starts from a condition and returns to the same condition, after having performed some work at the expenses of the heat extracted from a source at higher temperature than a sink. The theorem states that the ideally reversible engine produces work from heat if and only if the sink temperature is lower than the source's, $T_2<T_1$.  If $T_2>T_1$ work must be  supplied to the engine. If on the other hand $T_1=T_2$ no work can be extracted. Being in theory fully reversible and designed to have the maximum possible thermodynamic efficiency, the Carnot cycle can be run in the "forward" direction and "in reverse". When running in the opposite direction, the same amount of work performed in the forward cycle is returned to the source as heat, with zero net energy consumed and zero net work extracted. The practical problem with such an ideal situation is that the heat is extracted from the source, and transferred to the sink, while remaining at constant temperature, $T_1$ and $T_2$, apparently contradicting the experimental observation that $Q$ flows from hot to cold bodies. Furthermore, to move from $T_1$ to $T_2$ and back, a rigorously lossless transformation is required (adiabatic), which in practical terms means to proceed at infinitely slow rate.

\bigskip

 \addcontentsline{toc}{subsection}{Entropy is the name we give to our losses }

\noindent \textbf{Entropy is the name we give to our losses   (Clausius, 1856).} The Second Law of thermodynamics is quite different from other laws in physics, since (i) there are many different statements for the same law, and (ii) it is only a qualitative description, rather than a quantitative relationship between physical quantities.
Clausius wanted to put Carnot’s theorem on a more general basis, considering that heat exchanges between a body and a thermal bath are always \textit{not} reversible in the real world, and imply a loss of energy to the environment.
He introduced the notion of entropy, $S$, as the ratio between heat exchanged and working temperature, encompassing both reversible and irreversible transformations in the  single inequality:

\begin{equation}
Q \left( \frac{1}{T_{low}} - \frac{1}{T_{high}} \right) = \Delta S \ge 0 
\bigskip
\end{equation}

Hence, it is usually said that the Second Law of thermodynamics introduces the notion of a time arrow.
Here we already could start thinking of the analogy with the operations being carried out in a digital computer, accompanied by a waste of heat. The computer is in principle maintained at constant temperature, however it is an engine consuming energy to perform a computation, and its temperature would increase (in the absence of refrigeration and heat removal) at each operation performed. This energy goes into flow of electrons that move around the integrated circuits, capacitors, resistances, connecting wires. We can use Maxwell equations to deduce the amount of power accompanying the current. However, the fundamental operations that the computer is doing are creating and destroying information, by using this electrical current to flip the bit states in its memory from 0 to 1 and vice versa. Is there a link between the logical operations of creating and destroying information, and the energy required to physically run the computing machine? 

\bigskip

 \addcontentsline{toc}{subsection}{Statistical mechanics definition of entropy }

\noindent \textbf{Statistical mechanics definition of entropy (Boltzmann 1875).} In order to make such a link, we must at least be able to find a connection between the macroscopic world of thermodynamics, and the microscopic world in which electrons move and collide with other electrons and lattice vibrations (phonons). The connection between the macroscopic and microscopic degrees of freedom was attempted by Boltzmann, by introducing the notion of \textit{micro-state}, that is a definition of the instantaneous condition of the microscopic degrees of freedom (i.e., positions and momenta) that make up a macroscopically observable state. As it is immediately evident, a macroscopic state can be obtained in a variety of microscopic ways: the air molecules in a room continuously change their micro-state while the overall temperature and pressure remain constant. 

Boltzmann introduced the following microscopic expression for the entropy, interpreted as an extensive function that "counts" the number of micro-states of the system:

\begin{equation}
S = k_B \ln \Omega
\label{eqbolz}
\bigskip
\end{equation}

$\Omega$ is the number of microscopic states compatible with a given set of thermodynamical constraints $(T, P, V, N, …)$. $\Omega$ is a very difficult quantity to compute, or even to estimate, except some very simple cases, such as the perfect gas:

\begin{equation}
\Omega = \frac{1}{N!} \left( 2mE \right)^{3N/2} V^N
\label{sacktr}
\bigskip
\end{equation}

This statement is valid in the "microcanonical" statistical mechanics ensemble at constant-$\{N,V,E\}$.  For this experimental set up, all micro-states are equiprobable at equilibrium. 

By constrast,
for constant-$\{N,V,T\}$ conditions, that is the "canonical" ensemble, micro-states are not equiprobable, but are distributed according to the Boltzmann probability $\exp(–E/k_B T)$, and the energy $E$ is replaced by $k_B T$ in the definition of $\Omega$, since $T$ is now constant and all energy values $E$ are allowed. That is, energy can fluctuate. Fluctuating quantities are not usually considered in macroscopic thermodynamics, which deals with average values at equilibrium. Macroscopically we expect a system to have both a well-defined temperature and well-defined energy. When we look at the microscopic scale, for a given temperature the energy can fluctuate between different values. The macroscopic condition is recovered because energy fluctuations $\Delta E$ are proportional to the (square root of) specific heat, an intensive quantity proportional to the number of degrees of freedom $N$ of the system:

\begin{equation}
(\Delta E)^2 = k_B T^2 N c_V
\label{eqcv}
\bigskip
\end{equation}

Hence, when we calculate the relative importance of energy fluctuations with respect to the absolute value of the intensive quantity energy, it is $\Delta E/E \propto \sqrt{N} / N = 1/ \sqrt{N}$. In other words, in the limit of a macroscopic system $N\sim 10^{24}$ the energy is practically constant. Eq.(\ref{eqcv}) is an example of a fluctuation-dissipation relation, establishing a relationship between the thermal fluctuations of a physical quantity (energy) and another quantity (the specific heat) that describes its dissipation.

\bigskip

 \addcontentsline{toc}{subsection}{CMOS power dissipation: how big is a bit? }

\noindent \textbf{CMOS power dissipation: how big is a bit?}
Logical units in digital computers are made by combining a number of transistors, carved with high density in the silicon chip. When the transistor is in a given logical state its current consumption is negligible. All energy dissipation takes place during transitions between logical states, and the source of this dissipation is the need to charge or discharge the related capacitors.
The energy dissipated to charge/discharge one CMOS transistor has a well established form:

\begin{equation}
E_{switch} \simeq \alpha C_{node} V^2 \simeq 0.01 \cdot 10^{-12} \cdot (3)^2 \simeq 0.1 \text{\, pJ} \simeq \mathcal{O}(10^7) k_B T
\bigskip
\end{equation}

\noindent for a supply voltage of 3 V, $C_{node}$ being the lumped capacitance, and $\alpha$ a coefficient including the clock frequency.\cite{wilt2013} 
In the Xeon Broadwell-E5 (14-nm technology) about 7,2 million transistors arranged in about 1 million logic gates (making up CPU, memories, controllers, etc.), are packed in 456 mm$^2$.
Therefore, each transistor covers about 8$\times$8 $\mu$m$^2$, with a thickness of $\sim$0.2 mm, that is about 150 billion Si atoms.
So, each atom dissipates about $10^{-4} k_B T$ at each switching of the transistor. 

However, switching is a collective, statistically uncorrelated process: atoms follow quantum mechanics, currents follows Maxwell’s equations. A question arises: is there a link between the heat dissipation and the use/transfer/loss of information?

\bigskip

 \addcontentsline{toc}{subsection}{Thermal noise and random bit flips }

\noindent \textbf{Thermal noise and random bit flips.}
The flow of electrons in any current-conducting medium, for example across a resistor,  is affected by thermal fluctuations that entail a voltage fluctuation, with a spectrum usually assumed to be Gaussian with zero mean. The Johnson-Nyquist formula (originally derived on the basis of the equipartition law \cite{jon28,nyq28}) gives the thermal noise power density as the product of thermal energy by the bandwidth, $P=k_BT\Delta f$. To fix a number, a 1 kOhm resistor at room temperature with a bandwidth of 1 Hz generates a RMS noise of 4 nV. Although a capacitor is ideally a noiseless device, when combined with resistors it generates noise. Due to the fact that $\Delta f$ is inversely proportional to $\sqrt{RC}$, and given the steady reduction of the oxide layer with increasing transistor density, the overall result of CMOS miniaturization is an increase in the RMS width of the Gaussian voltage noise. 

Under such conditions, there is a finite probability that a spike in the voltage noise $V_n$ could pass, every now and then, the threshold voltage $V_{th}$ to flip the bit, in a random fluctuation. The average frequency by which such an event can occur can be estimated from the Rice formula,\cite{rice45} whose result in the approximation of white noise is:

\begin{equation}
\nu = \nu_0 \exp \left( - \frac{V_{th}^2}{2V_n^2} \right)
\bigskip
\end{equation}

Due to the steep dependence on the square of the $V_{th}/V_n$ ratio, the frequency of a random bit-flip is estimated about 35 million per hour at $V_{th}/V_n\simeq 5$, and drops to 2 in $10^{-9}$ per hour (that is, about 20 errors per hour in a 1-Gbyte RAM chip) at $V_{th}/V_n\simeq 10$.\cite{kish2002}
In recent years the threshold voltage $V_{th}$ has been constantly reduced, proportionally to the decrease in supply voltage, and in the most advanced CMOS circuits it could be of the order of $\sim$0.45 V.  Requiring the signal/noise voltage ratio to be at least $>$10, implies that the RMS thermal voltage fluctuation must be kept below about 35 mV to ensure a safe operation, or a maximum temperature of $T=eV_n/k_B\lesssim 150^o$C. Since leakage currents start affecting silicon electronics above $\sim$180$^o$C, random bit flipping is likely the main thermal limit to further decrease of voltage, the peak temperature of hot-spots in dense multiprocessor arrays being in the $\sim$100-120$^o$C (normal electronics is rated to function up to 85$^o$C, military electronics up to 125$^o$C, which actually seems a bit of a stretch, in view of the rapidly increasing bit-error rate).

\section{Information and thermodynamics: the demons of Leo Szilard}

 \addcontentsline{toc}{subsection}{“Information is physical” }

\noindent \textbf{“Information is physical” (R. Landauer, 1961).} One important achievement in the study of information processing has been to make the link with thermodynamics, with the understanding that manipulating information is inevitably accompanied by a certain minimum amount of heat generation. 
Computing, like all processes proceeding at a finite rate, must involve some dissipation. More fundamentally, there is a minimum heat generation per operation, independently on the actual rate of the process.
The binary logic devices of digital computers must have at least one degree of freedom associated with the information they carry, typically a logic port with more than one input and just one output mixes information from the input data, to present a value to the single degree of freedom of its output. As we will see below, devices with more input ports than output ports are inherently irreversible, in that the output does not allow to reconstruct the input information, Such devices exhibiting logical irreversibility are essential to classical computing. The important point is that  \textit{logical} irreversibility implies \textit{physical} irreversibility, which is accompanied by dissipative effects. The Boltzmann expression, such as Eq.(\ref{sacktr}), makes a link between entropy and the number of microstates available for a system at a given energy, showing that the larger is $\Omega$, the larger the distribution of possible configurations (in quantum terms, we could think of some analogy with mixed states). The dynamical equations, perfectly reversible at the level of individual degrees of freedom, become practically irreversible when the number of degrees of freedom gets very large. If we film two colliding balls and play the movie in reverse, it is impossible to tell the past from the future. If however we film a single ball hitting a triangle at snooker and play the same trick, it is immediately evident that the future is the one with more disorder at the end: the larger $\Omega$ brings more entropy, and less information about the dynamics of the individual trajectories. For a snooker with an Avogadro's number of balls, the information about past physical trajectories is irreversibly lost.

Rather than counting micro-states \textit{\`a la} Boltzmann, entropy can also be rewritten (Gibbs,\cite{jaynes65}) in terms of the absolute probability of each micro-state:

\begin{equation}
S = - k_B \sum_i p_i \ln p_i
\bigskip
\end{equation}

For the microcanonical ensemble in which all the $p_i =1/\Omega$, this writing is exactly the same as Boltzmann’s Eq.(\ref{eqbolz}) (which was actually put down in that form by Max Planck); for a distribution $p_i =\exp(-E_i/k_BT)$, instead, it easily shown that the canonical ensemble is obtained with constant $N,V$ and $T$.\cite{lebo20}

The notion of \textit{information entropy} was defined by Shannon,\cite{shann48} when he tried to quantify the “loss of information”:

\begin{equation}
H = - \sum_i p_i \ln p_i
\bigskip
\end{equation}

It can be viewed as the entropy change due to the presence/absence of information about a system, and it actually was Von Neumann to suggest Shannon the evident equivalence between his definition and Boltzmann's statistical mechanics formulation. But pushing the analogy even more forward, couldn’t it also be a measure of a \textit{heat loss} accompanying the exchange of information?

\bigskip

 \addcontentsline{toc}{subsection}{The Szilard engine }

\begin{figure}[h]
\centering
\includegraphics[width=\textwidth]{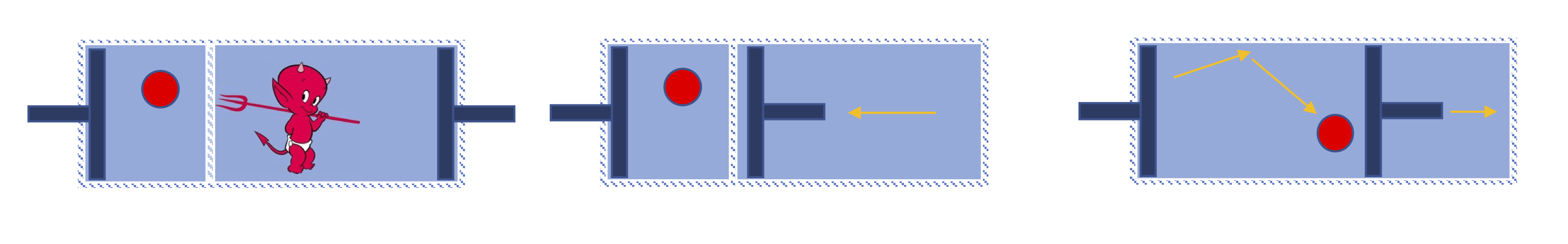}
\caption{The Szilard engine and its demon.}
\label{figszi}
\end{figure}

\noindent \textbf{The Szilard engine.} Instead of considering a gas made of a large number of particles (Carnot), consider just one single particle that is either on the left or on the right of a chamber equipped with two frictionless “pistons” and a “wall”. “Left” or “right” positions can be used to encode one bit of information (Figure \ref{figszi}). 
A “demon” who knows in which side of the box the particle is at time t=0, can spend this information (entropy) to: 

\begin{enumerate}
\item close the wall between the two halves of the box; then 
\item let the piston in the empty side move by doing zero work, until reaching the closing wall; and finally 
\item extract  useful work, by opening the wall and leaving the particle to expand back (isothermally) to its original equilibrium volume. 
\end{enumerate}

This thought experiment was designed in 1929 by Leo Szilard,\cite{szilard} to prove that possessing and using pure information has measurable thermodynamic consequences. Denoting by $p$ the probability that the particle is (for example) found on the left, the Shannon entropy for the 1-particle engine reads:

\begin{equation}
H[p] = -p \ln p - (1-p) \ln(1-p)
\bigskip
\end{equation}

If the demon has zero information, then: $p=(1-p)=1/2$, and $H[p]=-\ln 1/2 = \ln 2$. But if the demon has this 1 bit of information, $H$ must go to zero upon completing the cycle and:

\begin{equation}
\Delta H = H_{fin} – H_{in} = -\ln 2
\bigskip
\end{equation}

Using the single bit of information thus corresponds to a \textit{reduction} in the entropy of the system. The global system entropy is not decreased, but information-to-free-energy conversion is possible. After the particle is confined in one side of the box, the system is no longer in equilibrium: it appears that using information changed the system state without apparently changing the energy. Notably, the Szilard engine has been recently realized experimentally, by using Brownian particles,\cite{toyabe} or single electrons.\cite{koski}  

(To be fair, the demon should have indeed more than just one bit of information: he must firstly decide at what place to put the wall, and then control the piston's direction of motion. Therefore, at least 3 bits of information are required.) 

Let us then consider a "computer" with $N$  binary bits. In the initial (prepared) all-zero state, all the $p_i=1$ and :

\begin{equation}
H_{in} = -N \ln 1 = 0 
\bigskip
\end{equation}

After thermal equilibration, each bit has equal probability $p_i=1/2$ of being found in the state 0 or 1. This is to say that the initial information can be dispersed in any of the $N$  states, and the information entropy is:

\begin{equation}
H_{fin} = N \ln 2
\bigskip
\end{equation}

When we restore the initial state (that is, flush all voltages to ground, and reset all bits to 0) a minimum heat of:

\begin{equation}
\Delta Q = k_B T \Delta H = k_B T N \ln 2 
\bigskip
\end{equation}

\noindent is wasted. The RESET operation (erasure) is irreversible, and the wasted information turns to heat. The huge difference between the theoretical $\ln 2$ and the 
$\mathcal{O}(10^9) k_B T$ observed for the heat dissipation of a real transistor is due to the accessory circuits and wiring around each bit of information, but the lower limit of $\ln 2$ is incompressible.

To check this absolute limit, imagine that our computer goes instead into a defined final configuration, in which some bits have a higher probability of being in a given state, as could be in the result of some calculation. 
For example, $N/2$ have $p=3/4$ and $N/2$ have $p=1/4$ of being in state 1 (and all have $1-p$ probability of being in 0 ):

\begin{equation}
H_{fin} = \Delta H = - \tfrac{N}{2} \ln \tfrac{3}{4} - \tfrac{N}{2} \ln \tfrac{1}{4} = N \left( \ln 4 - \ln 3^{1/2} \right) > N \ln 2 
\bigskip
\end{equation}

It is easily shown that any choice of  $p_i$'s  different from  $1/N$  gives a larger entropy. The value of  $\ln 2$  appears therefore as an absolute lower bound for the heat dissipated by an operation destroying the information of a single bit.

\smallskip 

For quantum systems, the statistical state is described by the density matrix $\rho$. The probability to have a certain state $\ket{n}$ out of a complete basis, as the outcome of a measurement, is $p_n=\bra{n} \rho \ket{n}$. Therefore, the (Shannon) information entropy for such a measurement is:

\begin{equation}
S(\rho) = \sum_n \bra{n}\rho\ket{n} \ln \left(  \bra{n}\rho\ket{n}  \right)
\label{shann}
\bigskip
\end{equation}

By changing to the basis in which the density matrix is diagonal, the entropy assumes its minimum value, and is called the Von Neumann entropy:

\begin{equation}
S_{VN}(\rho) =-\Tr \{ \rho \ln \rho \}
\label{eqvonn}
\bigskip
\end{equation}

\noindent usually multiplied by the constant $k_B$ to give the entropy energy-like units. Evidently, the Von Neumann entropy can be identified with the information entropy only if we pretend to know beforehand in which basis $\rho$ is diagonal. For an equilibrium state with Hamiltonian $\mathcal{H}$, this is the canonical state, $\rho=\exp ( -\beta\mathcal{H} )$. A rigorous definition of entropy, however, should not assume any special \textit{a priori} basis. At least the (classical) uncertainty of the macroscopic measurement apparatus should be included, the quantum state entropy being written as a conditional probability $S(\rho|\mathcal{A})$ of obtaining a certain measurement outcome for a given measurement condition, and averaged over all the measurable results $\mathcal{A}$:\cite{stotland}

\begin{equation}
\langle S(\rho|\mathcal{A}) \rangle_{\mathcal{A}}  = S(\mathcal{A}) + \sum_{\mathcal{A}} P(\mathcal{A}) S(\rho|\mathcal{A})
\bigskip
\end{equation}

\noindent where $\langle ... \rangle_\mathcal{A}$ indicates ensemble averaging, $P(\mathcal{A})$ is the probability of finding the macroscopic measurement apparatus in the condition $\mathcal{A}$, and $S(\mathcal{A})$ the corresponding classical entropy.


\bigskip

 \addcontentsline{toc}{subsection}{Landauer’s Principle (classical) }

\noindent \textbf{Landauer’s Principle (classical).} Logical irreversibility is the act of processing an information in which the output does not uniquely permit to retrieve the inputs. Now, the link between Gibbs' and Shannon's definitions of entropy is a purely mathematical one: by dealing with two very different situations, different variables and different processes, they arrive at two definitions of a measurable quantity that formally read identical.  A plausible deduction is that these should be therefore the same quantity. Landauer,\cite{landauer61} and later Bennett,\cite{bennett82} tried to put the equivalence on more physical grounds. Their idea was that information at its most basic stage is a distribution of 0's and 1's physically entrusted to a set of bistable systems described by a bistable potential; then, the (classical) thermodynamics of each two-state system automatically associates the processing of information with the thermodynamics laws that those physical systems ought to follow. As a result, Shannon, Gibbs and Clausius' entropies \textit{must} describe the same thing, and logical irreversibility \textit{must} imply thermodynamic irreversibility in the sense of the Second Law, that is, increase of entropy.\cite{ladyman}

The only nontrivial reversible operation a classical computer can perform on a single bit is the NOT operation, with one input and one output whose values are strictly defined. By contrast, the operations AND, NAND, OR and XOR are all irreversible, since they have more than one input and just one output. 
Hence, from the output of these logic gates we cannot reconstruct the input: information is irreversibly lost. 
However, in a quantum computer irreversible operations may be - at least ideally - avoided, for example by saving the entire history of the process, or by replacing the irreversible gates by more complex but reversible gates, e.g. using a Toffoli gate instead of the AND. It seemed therefore that information processing at the quantum level should have no intrinsic thermodynamic cost, as firstly Bennett\cite{bennett82}  and then Feynman observed.\cite{feynman85}  

But the operation of erasing a bit of information, instead, has two possible states (0 or 1) being mapped to a single definite state of 0, so it must entail a loss of entropy since the value 0 has now $p$=1, for both a classical or a quantum computer. A reformulation of the Second Law, Landauer’s principle states that the entropy decrease of the information-carrying degrees of freedom must always be compensated by an equal (or greater) entropy increase in the environment.

\bigskip

 \addcontentsline{toc}{subsection}{Quantum computation is microscopically reversible }

\noindent \textbf{Quantum computation is microscopically reversible.}
Qubits are defined as two-state quantum systems, described by a state vector in a 2-dimensional Hilbert space, spanning a closed surface with conserved norm (i.e., the Bloch sphere). 
A standard basis is defined by two vectors $\ket{0}$ and $\ket{1}$, conventionally aligned with the positive and negative direction of the z-axis.

In principle, a quantum gate performs rotations in the Bloch sphere of one or more qubits onto which it is applied. Therefore, any quantum gate is a 
unitary operator:

\begin{equation}
UU^{\dagger} = U^{\dagger}U=\textbf{I}
\bigskip
\end{equation}

\noindent \textbf{I} being the identity operator. Such operation  conserves the norm of the quantum state, and is perfectly time-reversible.\cite{bennett82,feynman85}  Upon application of any sequence of quantum gates, state vectors span the whole surface of the Bloch sphere. Being unitary, qubit rotations (in principle) do not generate any heat.

A \textit{pure} quantum state is one that cannot be written as a probabilistic mixture of other quantum states. Pure states can also result from the superposition of other pure quantum states (entanglement). A density matrix, $\rho=\ket{\psi}\bra{\psi}$, can be used to represent both pure ($\rho^2=\rho$) and mixed ($\rho^2\ne\rho$)  states. 
Let’s look for example at two states in the 2-dim Hilbert space of a qubit, $\ket{\psi_1}=\left(\begin{smallmatrix}1 \\ 0\end{smallmatrix}\right)$ and $\ket{\psi_2}=\left(\begin{smallmatrix}0 \\ 1\end{smallmatrix}\right)$. Then, for a mixed state with equal probabilities $p_i=\tfrac{1}{2}$ we have:

\begin{equation}
\rho = \sum_i p_i \ket{\psi_i} \bra{\psi_i}=\frac{1}{2} \left( \begin{matrix} 1 & 0 \\ 0 & 1 \end{matrix} \right)
\bigskip
\end{equation}

On the other hand, for a pure state with equal amplitudes, $\ket{\psi}=\tfrac{1}{\sqrt{2}}(\ket{\psi_1}\bra{\psi_2}$, the density matrix is:

\begin{equation}
\rho = \ket{\psi} \bra{\psi}=\frac{1}{2} \left( \begin{matrix} 1 & 1 \\ 1 & 1 \end{matrix} \right)
\bigskip
\end{equation}

In the Bloch sphere representation of a qubit, each point on the unit sphere stands  for a pure state. 
The arbitrary state for a qubit can be written as a linear combination of the Pauli matrices $(\hat{\sigma}_x,\hat{\sigma}_y,\hat{\sigma}_z)$, with three real numbers $(r_x,r_y,r_z)$ as the coordinates of a point in the sphere:

\begin{equation}
\rho = \frac{1}{2} \left( \textbf{I} + r_x \hat{\sigma}_x + r_y \hat{\sigma}_y + r_z \hat{\sigma}_z \right)
\label{eqpaul}
\bigskip
\end{equation}

Points for which $r^2_x + r^2_y + r^2_z = 1$ lie on the surface, and represent pure states of any superposition of $\ket{\psi_1}=\left(\begin{smallmatrix}1 \\ 0\end{smallmatrix}\right)$ and $\ket{\psi_2}=\left(\begin{smallmatrix}0 \\ 1\end{smallmatrix}\right)$. Any other combination of $r^2_x + r^2_y + r^2_z < 1$ lies in the interior of the sphere, and represents thermally-mixed states.

\smallskip 

How can we get thermally mixed states starting from pure states, and perform unitary transformations that should not generate any heat loss?

\bigskip

 \addcontentsline{toc}{subsection}{Pure states vs. mixed states: quantum entropy }

\noindent \textbf{Pure states vs. mixed states: quantum entropy.} 
The time evolution of a pure state starting from $\rho^0$  at time $t=0$ under the action of a unitary operator $\hat{U}(t)=exp(-i\mathcal{H}t/\hbar)$ is obtained from the Von Neumann equation:

\begin{equation}
\frac{d\rho}{dt} = - \frac{i}{\hbar} \left[ \mathcal{H},\rho \right]
\label{timevol}
\bigskip
\end{equation}

\noindent (that is, the quantum-equivalent of Liouville's equation) as : 

\begin{equation}
\rho^t = U(t) \rho^0 U^{\dagger}(t)
\bigskip
\end{equation}

For a time-independent Hamiltonian it is easily shown that the density matrix elements evolve as: 

\begin{equation}
\rho_{nm}(t)=e^{-i\omega_{nm}(t-t_0)}\rho_{nm}(t_0)
\bigskip
\end{equation}

The intrinsic dynamics generated by this time evolution is unitary, i.e. the diagonal density $\rho_{nn}$ is conserved in time, and the coherent superpositions oscillate at the frequencies $\omega_{nm}$. The Von Neumann entropy, Eq.(\ref{eqvonn}),
is as well invariant under unitary dynamics (in fact, for pure states this $S(\rho)$  is just zero). This means that entropy generation by irreversibility cannot be a result of the intrinsic quantum dynamics.  
It can only result from changes in time of the statistical description of the interaction with an external system, which turns pure states into mixed states.

The density matrix of a mixed state can be defined on the basis of all the pure states $\ket{\psi_i}$  as :

\begin{equation}
\rho = \sum_i p_i \ket{\psi_i} \bra{\psi_i}
\bigskip
\end{equation}

\noindent and the Von Neumann quantum entropy of the mixed state, by extension, is obtained as :

\begin{equation}
S_{VN}(\rho) = -k_B \Tr \left\{ \rho \ln \rho \right\} = - k_B \sum_i p_i \ln p_i
\bigskip
\end{equation}

For two entangled subsystems $A$ and $B$ (for example two qubits, or an atom and an external field) a quantity of interest is the Araki-Lieb inequality:\cite{arakilieb}

\begin{equation}
| S_A - S_B | \leq S_{AB} \leq S_A + S_B 
\label{araki}
\bigskip
\end{equation}

 For a pure state, the partial trace tells that the entropy is equal for the two subsystems $S(\psi)=-\Tr{\rho_A \ln \rho_A} =-\Tr{\rho_B \ln \rho_B}$. The inequality (\ref{araki}) gives the same result, because the total wavefunction is also a pure state, therefore $S_{AB}=0$, which implies $S_A=S_B$. This may be very useful e.g. for the case of an spin-1/2 atom interacting with an external field: while the entropy of a two-state system is easy to calculate, the entropy of the field could be much more difficult to obtain. It has been recently demonstrated that the same Araki-Lieb inequality can be extended to mixed states.\cite{anaya}

\section{Quantum thermodynamics is not what you think}

It is both interesting and funny to think that, to some extent, quantum mechanics was born out of thermodynamic considerations. The energy quantum was introduced in 1900 by Max Planck as a last resort in the search for an explanation of the experimental plots of thermal blackbody radiation. Five years later, Einstein introduced the first germ of the idea of quantization of the electromagnetic field, on the basis of thermodynamic equilibrium of the blackbody "resonators". And Einstein again, in 1916, explained the relation between stimulated emission and radiation absorption using as well thermodynamic equilibrium arguments, in a seminal paper that represents the theoretical birthdate of lasers.\cite{einstein}

\bigskip

 \addcontentsline{toc}{subsection}{Temperature? }

\noindent \textbf{Temperature?}
Temperature is at the heart of both classical thermodynamics and statistical mechanics, and yet it is a rather difficult notion to put on firm grounds. The schoolbook definition of temperature as "average kinetic energy of the system" makes little sense upon closer inspection, unless only the translational kinetic energy is considered: the amount of energy to increase temperature by 1 degree is different for a monoatomic vs. a diatomic gas. Kelvin's definition of absolute temperature focused on the heat exchanges between thermal baths, in the style of Carnot (who in his time did not have the concepts of heat and entropy, and spoke generally of "caloric"), defining the \textit{ratio} of two temperatures as being equal to the ratio of the exchanged heat between two bodies. The more formal definition (Gibbs) looks at the change in entropy as a function of internal energy, at constant-$\{N,V\}$:

\begin{equation}
\frac{1}{T} = \frac{\partial S}{\partial E} \bigg|_{N,V}
\label{eqtmp}
\bigskip
\end{equation}

\noindent and defines temperature as an intensive quantity, the ratio between the differentials of two extensive variables. 

As we saw in the previous section, defining entropy rigorously for quantum systems with discrete energy levels is still problematic, and this holds even more true for the notion of temperature. Temperature is a property of the aggregate system, not of each single particle, and is properly defined only for systems at equilibrium. Instead, open quantum systems are often found in non-equilibrium states, strongly coupled and correlated with the environment. Temperature is classically an intensive variable, that is, a physical quantity that can be measured locally and is the same throughout the system; however, for systems with strong interactions and a small number of degrees of freedom, locality is lost and some equivalent of temperature could no longer be intensive.\cite{hartmann,ferraro} In standard quantum statistical mechanics, temperature is treated just as a parameter in the wavefunction, and does not have an operator associated (you usually see it just as the $\beta=1/k_BT$ relief at the denominator of the Fermi-Dirac or Bose-Einstein energy exponentials).

For quantum systems with sufficiently close-spaced energy levels, it is customary to use Von Neumann's definition, Eq.(\ref{eqvonn}) (which strictly speaking refers to information and not to heat exchanges).\cite{vallejo20} 
The reduced density matrix of a qubit in a random point of the Bloch sphere (see Eq.(\ref{eqpaul})) is then:

\begin{equation}
\rho_r = \frac{1}{2} \left( 1 + \vec{r}\cdot\vec{\sigma}\right) =  \frac{1}{2} \left( \begin{matrix} 1+r_z & r_x-i r_y \\  r_x+i r_y & 1-r_z \end{matrix} \right) 
\bigskip
\end{equation}

\noindent $\vec{\sigma}$ being the vector with components the Pauli matrices, and the entropy for the modulus $r=|\vec{r}|$ is:

\begin{equation}
\frac{S_r}{k_B} = -\left(\frac{1+r}{2}\right)  \ln \left(\frac{1+r}{2}\right)  -\left(\frac{1-r}{2}\right)  \ln \left(\frac{1-r}{2}\right)
\bigskip
\end{equation}

Let us imagine for the sake of simplicity a spin-qubit $\mu$ in a magnetic field $\vec{B}$, with Hamiltonian $\mathcal{H}=-2\mu(\vec{r}\cdot\vec{B})$. The internal energy in the density-matrix formalism is defined $E=\langle \mathcal{H} \rangle = \Tr \{\rho_r \mathcal{H}\}$, from which a quantum equivalent "temperature" follows by formally applying Eq.(\ref{eqtmp}):

\begin{equation}
 T = \frac{1}{k_B} \frac{\mu r}{ B_{||} \tanh^{-1}r}  
\bigskip
\end{equation}

\noindent $B_{||}$ indicating the component of $\vec{B}$ projected on $\vec{r}$. It can be seen that at this level, thermodynamic
properties, and in particular the temperature, are function only of $\vec{r}$. Note that for pure states on the Bloch surface, the Von Neumann entropy is zero and such a definition of temperature also goes to zero (with the puzzling consequence that one could get to absolute zero by using a finite quantity of energy, thus contradicting the Third Law of thermodynamics). On the other hand, temperature is intended an average quantity that is applicable only to a system with a large number of degrees of freedom, and in contact with a thermal bath, that is in a mixed state.

\smallskip 

As a next step, let us consider an isolated system of $2N$ spin qubits, initially prepared in an eigenstate with total energy $E$. At a given temperature, a subset $M$ of spins is excited. For a weak coupling, it must be $M \ll 2N$. This is a \textit{microcanonical} ensemble that at thermal equilibrium must equally share the total energy between all its degrees of freedom $\Omega$. To have a density operator that is diagonal in any base, we must require that the wavefunction is an incoherent superposition of all states with constant energy $E$ and random phases $\phi_j\in[0,2\pi]$:\cite{ghonge} 

\begin{equation}
 \ket{\psi} = \frac{1}{\sqrt{R}} \sum_{j=1}^R e^{i\phi_j} \ket{\psi_M^{(j)}},\,\,\,\,\,\,\,\,\,\,\,R=\left(  \begin{matrix}   2N \\ M     \end{matrix} \right)
\bigskip
\end{equation}

$R$ is the number of states with $M$ thermally excited spins, and $\ket{\psi_M^{(j)}}$ is the $j$-th wavefunction of such ensemble. For example, with $N$=4 and $M$=2, it is $R$=28 and $\ket{\psi_2^{(1)}}$=$\ket{11000000}$, $\ket{\psi_2^{(2)}}$=$\ket{10100000}$, ... $\ket{\psi_2^{(28)}}$=$\ket{00000011}$. For a given excitation $B$, the temperature is a function of $R$:

\begin{equation}
 T_R = \frac{1}{k_B} \frac{2\mu B}{ \ln (2N/M-1)}  
\bigskip
\end{equation}

An increasing temperature corresponds to an increasing fraction, $M\rightarrow N$, of spins excited; at the opposite, $T\rightarrow 0$ when the system tends to perfect paramagnetic alignment.
Anyway, the definition of temperature in the quantum regime still remains a subject of fundamental and "heated" discussions.\cite{kosloff,hartmann,ghonge,bartosik}

\bigskip

 \addcontentsline{toc}{subsection}{Quantum Carnot }

\noindent \textbf{Quantum Carnot.} 
Classically, the Carnot engine consists of two sets of alternating adiabatic strokes and isothermal strokes. A priori, one may argue that the laws of thermodynamics (with an exception for the First) are defined just for macroscopic systems described by statistical averages, and hence, the question of their validity for microscopic systems consisting of a few particles, or qubits, may itself appear meaningless. However, already in 1959 \cite{scovil59} Scovil demonstrated that the working of a quantum three-level maser coupled to two thermal reservoirs resembles that of a heat engine, with an efficiency upper-bounded by the Carnot limit. A quantum analogue of the Carnot engine consists of a working fluid, which could be a particle in a box,\cite{bender00} qubits of various kind,\cite{geva92} multiple-level atoms,\cite{quan07} or harmonic oscillators.\cite{lin03} For the simplest case of a three-level system, the quantum working fluid is the spectrum of energy levels $E_1< E_2 < E_3$; the high-temperature bath can excite transitions $\hbar\omega_h=|E_1-E_3|$, the low-temperature sink induces transitions  $\hbar\omega_c=|E_1-E_2|$; and a radiation field is tuned resonantly at the frequency  $\hbar\omega_r=|E_2-E_3|$. At equilibrium, for each excitation $\hbar\omega_h$ the system loses an energy $\hbar\omega_c$ to the cold sink, and $\hbar\omega_r$ to the radiation, so that the population ratios $n_1/n_3=\exp(\hbar\omega_h/k_BT_h)$ and $n_1/n_2=\exp(\hbar\omega_c/k_BT_c)$ are maintained steady. The energy exchanged with the two thermal baths can be thought of as “heat” (positive or negative), while the energy exchanged with the radiation field can be identified with “work” extracted from the quantum system (the radiation plays the same role as Carnot’s “piston”). This identification of work and heat implies the energy relation  $\hbar\omega_h= \hbar\omega_c+\hbar\omega_r$,  which is the analog of entropy conservation for a reversible cycle. (Reversibility here is within the limit of statistical equilibrium among all the excitations.) 

A remarkable result appears when the efficiency of this “thermodynamic” system is considered. The quantum system can work as an engine when a population inversion is realized between the levels 2 and 3, $n_2>n_3$, which leads to a condition: 

\begin{equation}
\frac{n_2}{n_3}=\frac{n_2}{n_1} \frac{n_1}{n_3}>1 = \exp \left( \frac{\hbar\omega_c}{k_BT_c} –\frac{\hbar\omega_h}{k_BT_h} \right) \ge 1
\bigskip
\end{equation}

The efficiency is, as usual, the ratio of the work extracted to the heat supplied by the hot reservoir, 

\begin{equation}
\eta = \frac{\hbar\omega_r}{\hbar\omega_h} = 1 - \frac{\hbar\omega_r}{\hbar\omega_h} 
\bigskip
\end{equation}

\noindent which - thanks to the previous inequality - gives Carnot’s limit $\eta \leq 1 - T_c/T_h$.

The proof of existence of Carnot’s limit (a manifestation of the Second law of thermodynamics) at quantum length scales establishes a strong case for the emergence of thermodynamic laws at the most fundamental level.
Quantum cyclic processes, although different in many ways from a Carnot cycle, still have important features in common with it.
Most importantly, however, it has been shown that a quantum engine could exceed the capabilities of the Carnot cycle, in that it can operate between reservoirs of positive and negative temperatures.\cite{geusic67} 

\begin{figure}[h]
\centering
\includegraphics[width=0.75\textwidth]{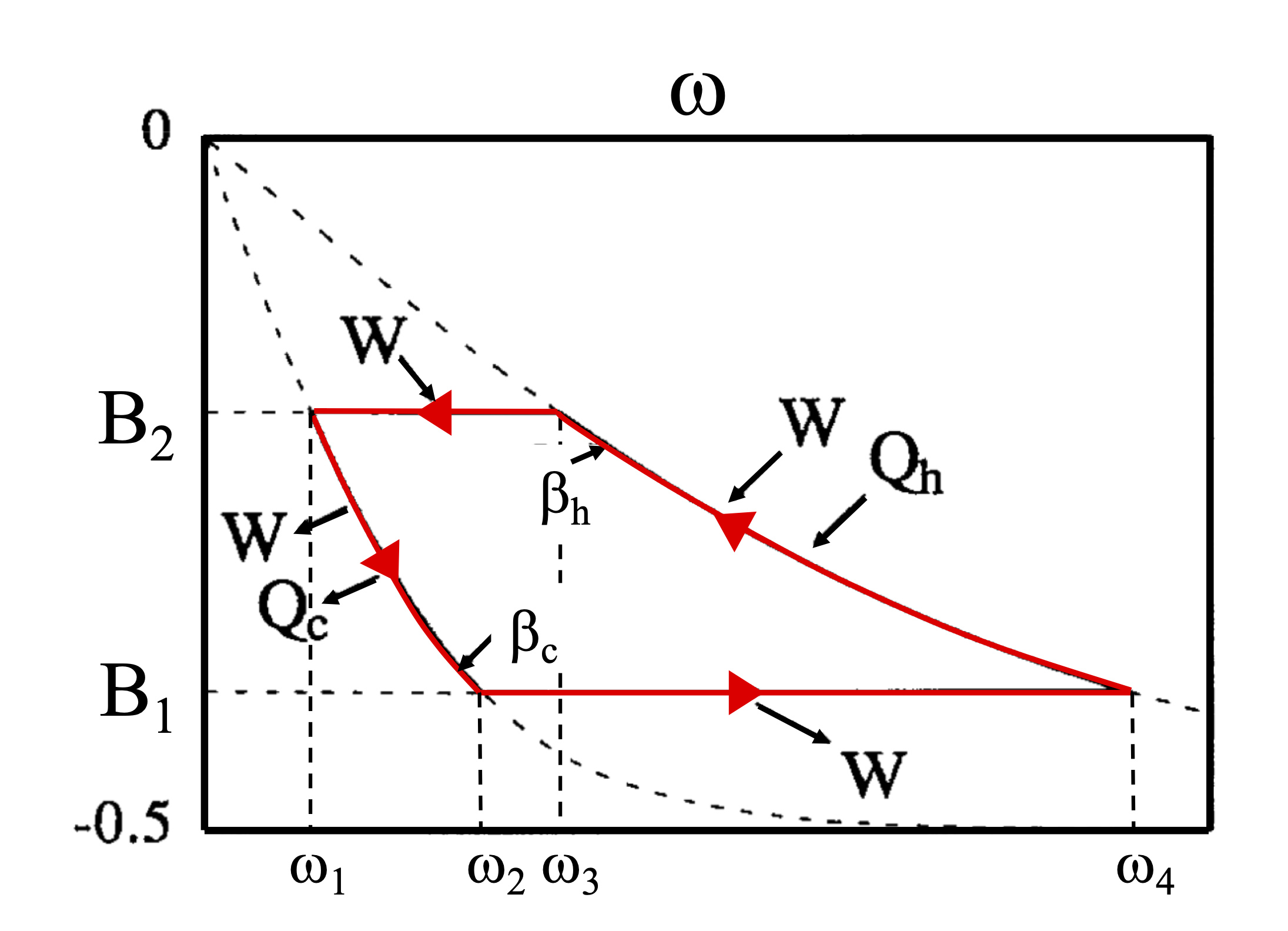}
\caption{ A reversible Carnot cycle depicted in the space of the normalized magnetic field $\omega$ and the magnetization $B$. The horizontal lines represent adiabats wherein the engine is uncoupled from the heat baths at inverse temperatures $\beta_{h,c}=1/k_BT_{h,c}$, and the magnetic field is changed between two values; the two horizontal strokes represent changing the magnetisation $\omega$ while the qubits are connected to a heat bath at constant temperature. Black arrows indicate the direction of heat and work from/to the spin-qubit system. (Adapted w. perm. from Ref.\cite{geva92}). 
}
\label{figspin}
\end{figure}

The classical definition Eq.(\ref{eqtmp}) of temperature as the variation of entropy with energy, allows in theory a negative value of temperature if for some system the entropy does not increase, but rather decreases upon increasing energy. The conditions by which this could happen were first identified by Onsager,\cite{onsa49} and more precisely stated by Ramsey,\cite{rams56} as far back as 1956. The simplest example is a 1D chain of 1/2-spins of non-interacting qubits with gyromagnetic constant $\gamma$, in a magnetic field $\omega(t)=-\gamma B_z(t)$.\cite{geva92}  The time-dependent Hamiltonian is simply $\mathcal{H}(t)=\hbar\omega(t)\sigma_z/2$, coupled to two baths at temperatures $T_h$ and $T_c$ (Figure \ref{figspin}). During the adiabatic expansion, $\omega_2 \rightarrow \omega_4$, and compression, $\omega_3 \rightarrow \omega_1$, work is done by, or on the spins, but entropy is constant; in the cold, $\omega_1 \rightarrow \omega_2$, and hot isotherm, $\omega_4 \rightarrow \omega_3$, both heat and work are transferred to the cold bath, or removed from the hot bath, while entropy, respectively, decreases or increases. The expectation value of the Hamiltonian is obtained as:

\begin{equation}
\frac{d\langle\mathcal{H}(t)\rangle}{dt} = \frac{1}{2} \left( \frac{d\omega}{dt} \langle \sigma_z \rangle + \omega \frac{d\langle \sigma_z \rangle}{dt} \right)
\bigskip
\end{equation}

The two terms on the RHS are to be identified, respectively, with the average work, $\langle \delta W \rangle = \langle \sigma_z \rangle \delta\omega/2$, and average heat, $\langle \delta Q \rangle = \omega \delta \langle \sigma_z \rangle /2$, exchanged, in the analog of the derivation of the First Law (see Eq.(\ref{eqweak}) below).

Now, consider the extreme situation in which the total magnetic energy of the spin chain increases continuously from the lowest state with all spins "down", up to the highest state with all spins "up". Both the initial and final states have just one microstate available, $\Omega$=1 $p_1$=1, therefore their Von Neumann entropy is zero. While the magnetic field spans between these two extremes, the entropy first increases from 0, goes to its maximum when the spins are on average half-up and half-down, and then decreases again going back to 0. Correspondingly, the temperature goes from zero to plus-infinity at the entropy maximum, then jumps to negative-infinity (because entropy and energy have opposite-sign first derivatives) and goes back to zero always from negative values.

As Ramsey pointed out,\cite{rams56} the number of physical systems actually capable of assuming a negative temperature is limited to systems with a finite number
of energy levels, and sufficient thermal insulation from positive-temperature reservoirs. In the real-world, atomic or nuclear spin systems have other degrees of freedom; if the coupling between the spins and other degrees of freedom is much weaker than the strong coupling between spins, we can talk about a "spin-temperature" separately from the temperature of the atoms, or lattice as a whole. It is interesting to note that if one can realize a system simultaneously coupled to a positive and a negative thermal bath,  Carnot's efficiency $(1-T_1/T_2)$ could indeed assume values larger than 1.\cite{geusic67} 

\smallskip

By using quantum mechanical states as the heat exchanger "fluid", even a single atom can turn into a Carnot engine, as shown by Singer's group in Mainz.\cite{rossnag1} Sandwiched between an electric field representing the hot reservoir and a laser cooling beam representing the cold reservoir, a single $^{40}$Ca$^+$ ion is caught in a funnel-shaped, magnetic quadrupole linear trap, with frequency $\omega_r$. The "temperature" of the ion quantum state is determined by the radial spreading of the wavefunction, approximately Gaussian with a width $\sigma(T)=(k_BT/m\omega_r^2)^{1/2}$. The cooling laser is always on, while the electric field switched on and off, thereby making the ion temperature oscillate between "cold" and "hot"; by sweeping the trap frequency between the extremes $\pm \omega_M$, a thermodynamic cycle is performed and work is extracted by the axial force generated by the movement of the trapped ion. Compared to the equivalent Carnot cycle, the efficiency is extremely small, of the order of 0.003, however the result of a single ion performing as a reversible and essentially frictionless quantum engine is nothing short of amazing.

The same group had previously demonstrated (but not experimentally realized) an example of a Otto cycle for a time-dependent oscillator coupled to a "squeezed" thermal reservoir, which could have a theoretical efficiency above Carnot's limit and approaching unity.\cite{rossnag2} There, the squeezing (a common concept in quantum optics\cite{breiten97}) refers to the particular construction of the quantum states of the thermal bath, in which the thermal noise is distributed differently among the degrees of freedom; for example, in a harmonic oscillator the noise can be concentrated in the phase but not in the amplitude.\cite{breiten97,esteve08}

\bigskip

 \addcontentsline{toc}{subsection}{Thermal vs. quantum fluctuations }

\noindent \textbf{Thermal vs. quantum fluctuations.}
Thermodynamics is a macroscopic effective picture of thermal processes, not concerned with microscopic details, but only dealing with average quantities such as temperature, work, dissipated heat. 
This approach is valid for a macroscopic number of particles, but starts losing accuracy as the system size decreases to a small number of degrees of freedom, and thermal fluctuations of the average quantities become more relevant than the averages themselves. 
Compared  with  macroscopic  thermodynamics,  fluctuations   play  a  much more  important  role  in  small systems. However, the presence of fluctuations does not mean that we cannot characterize quantum systems thermodynamically; on the contrary, fluctuations typically contain important additional  thermodynamic  and  energetic  information  that  is  usually  lost as noise in  the infinite-system  limit. 

Stochastic thermodynamics picks up where the macroscopic description starts to fail, and gives insight into the fluctuations of thermodynamic quantities. It also moves beyond the equilibrium situations associated with thermodynamics, and can describe the behavior of systems that are out of equilibrium. 
Considerations stemming from fluctuation theorems \cite{evans1993,jarz1997,crooks1999}  are vital when considering nanoscale devices, or biological protein machines, for which experiments confirmed the theoretical predictions of local violation of the Second Law.\cite{evans2004} 
Such "theorems" (in fact, they should be better called "relations", since they do not stem from a rigorous derivation from a set of axioms) state at different levels that for dynamical systems far from equilibrium there exists a physically meaningful, real-valued variable $\Omega_t$, extensive both in space and time, whose positive values are exponentially more probable than the negative ones, or:

\begin{equation}
\frac { P(+\Omega_t) }{ P(-\Omega_t)} =e^{ \Omega_t }
 \bigskip
\end{equation}

 In practice, this variable is easily identified with the entropy production, extensive and increasing with time. 
What such fluctuation relations state, therefore, is that the Second Law probabilistically holds for a macroscopic system observed over macroscopic times. And it can be "violated" (that is, entropy flows in the reverse direction) is the system is sufficiently small and/or the observation time is sufficiently short.
In particular, according to Crook's fluctuation relation,\cite{crooks1999} for a transformation between two microscopic states $A$ and $B$ separated by a free energy $\Delta F$, the thermodynamic work $W$ is a fluctuating quantity, and is therefore given by a probability distribution of values. For ideally reversible transformations, the work distributions in the time-forward or backward direction cross at the value $W=\Delta F$, as clearly demonstrated by optical tweezer experiments on the cyclic folding and unfolding of RNA fragments.\cite{felix2005}

\smallskip

At the even smaller scale, however, fluctuations are no longer just thermal, but quantum-mechanical in origin. In the regime in which quantum phenomena are manifest, that is, very low temperatures and sizes smaller than the De Broglie wavelength, Heisenberg's uncertainty relations become the relevant source of noise in the form of localized, temporary random changes of the system energy, for a (very) short time.
Then, many questions arise when applying such concepts to qubits. For example:
\begin{itemize}
\item What becomes of thermodynamic equilibrium, for time-reversible, unitary transformations? 
\item What is the meaning of thermalization in the presence of quantum integrals of motion? 
\item How to define and/or measure thermodynamic quantities for quantum systems? 
\item How entanglement is connected with the information entropy?
\item and more…
\end{itemize}

The fluctuations we are after for a quantum system in contact with a heat bath are not strictly thermal ones. Rather, they are represented by combinations of (1) the possible changes in the distributions of the energy levels (that is, a change of the Hamiltonian), 
and/or (2) changes of their occupation numbers (that is, entropy). In both cases, the result is a degradation of the quantum state, i.e. a loss of coherence. The pure state turns into a mixed state.

In the approximation of weak coupling between the quantum system and the thermal bath, the equilibrium density 
tends to a Gibbs state ($\beta=1/k_b T)$:

\begin{equation}
\rho^{eq} = \frac {\exp(-\beta \mathcal{H})} {\mathcal{Z}}
\bigskip
\end{equation}

\noindent with as usual $\mathcal{Z}=\Tr \{ \exp(-\beta \mathcal{H})\}$ the system's partition function. The average internal energy is $E(\rho)=\Tr\{\mathcal{H}\rho\}$, the entropy $S=-\Tr\{\rho \ln \rho\}$, and the free energy is obtained as $F = E(\rho) - TS(\rho) = - \ln \mathcal{Z} / \beta$. Hence, a "weak" quantum-equivalent of the First Law can be written:

\begin{equation}
dE = \delta Q + \delta W = \Tr \{ (d\rho^{eq}) \mathcal{H} \}  + \Tr \{ \rho^{eq} (d\mathcal{H}) \} 
\label{eqweak}
\bigskip
\end{equation}

The first term on the right-hand side, containing the differential of the equilibrium density, is relative to a variation of occupation numbers of the quantum eigenstates, and is therefore assimilated to a form of thermodynamic entropy analogous to the $\delta Q$ of classical thermodynamics; the second term, in turn, containing the differential of the system Hamiltonian, corresponds to a change in the structure of the energy levels, as it could derive by a change in the system mechanics, and can be assimilated to a work $\delta W$ done on, or by, the quantum system.

\bigskip 

 \addcontentsline{toc}{subsection}{Work and heat are not quantum-mechanical observables }

\noindent \textbf{Work and heat are not quantum-mechanical observables.}  Both quantities are dependent on the process path $\lambda$  (and thus are non-exact differentials, like in classical thermodynamics), which means they do not correspond to quantum-mechanical observables, i.e. there is no Hermitian operator $\hat{q}$ or $\hat{w}$ such that $Q = \Tr \{ \rho \hat{q} \}$ and $W = \Tr \{ \rho \hat{w} \}$.
The intuitive, simplistic reasoning behind such a statement is that the final-state Hamiltonian at $t$=$\tau$ does not necessarily commute with the initial Hamiltonian at $t$=0, i.e.

\begin{equation}
[ \mathcal{H}(\lambda^t),\mathcal{H}(\lambda^{\tau}) ] \ne 0
\bigskip
\end{equation}

\noindent for some (or all) times $0<t<\tau$.

A different definition of \textit{quantum work} (w/r to Eq.(\ref{eqweak})) can be given as the difference between eigenvalues of the “instantaneous” Hamiltonian at the beginning and end of the path $\lambda$:

\begin{equation}
W = \left( \epsilon^{\lambda_{\tau}}_m - \epsilon^{\lambda_{0}}_n    \right)
\label{pathw}
\bigskip
\end{equation}

Here quantum work is a random variable distributed as $p(W;\lambda)$, and is given by a time-ordered correlation function as a path-dependent quantity. On this basis, the quantum-equivalent of the fluctuation theorems can also be recovered:\cite{talkner2007,talkner2011}

\begin{equation}
\frac{p(W;\lambda)} {p(-W;\lambda)} = e^{\beta (W - \Delta F) }\,\,\,\,\,\,\,\,\,\,\text{(Crooks)}
\bigskip
\end{equation}

\begin{equation}
\langle e^{\beta W } \rangle_{\lambda} = e^{-\beta \Delta F }\,\,\,\,\,\,\,\,\,\,\,\,\,\,\,\,\,\,\text{(Jarzinsky)}
\bigskip
\end{equation}

However, for a quantum system entangled with its environment the interaction energies are not weak, in fact they will quickly degrade the pure state into a mixed one, in a time of the order of the coherence time. Identification of "heat" and "work" with the variation of the system's characteristics $(d\rho,d\mathcal{H})$ is no longer enough. During isothermal quasi-static processes, part of the free energy exchanged with the environment represents an "energetic price" to pay, in order to preserve the coherence and quantum correlations in the system.
Denoting a non-Gibbsian, \textit{coherent} and \textit{correlated} state as $\rho^{cc}$, the extended entropy $S_e$ can be written as :

\begin{equation}\begin{split}
S_e &= - \Tr\{\rho^{cc} \ln{\rho^{cc}} \} = \\
\\
       &= - \Tr\{\rho^{cc} \ln{\rho^{cc}} \} + \left[ \Tr\{\rho^{cc} \ln{\rho^{eq}} \} - \Tr\{\rho^{cc} \ln{\rho^{eq}} \} \right] = \\
\\
      &= \beta [ E - (F + TS(\rho^{cc} || \rho^{eq}) ] = \beta [ E - \mathcal{F} ]
\bigskip
\end{split}\end{equation}

\noindent (note that the last term in [...] in the second line is zero). 
$S(\rho^{cc} || \rho^{eq}) =  \Tr\{\rho^{cc} ( \ln{\rho^{cc}} -\ln{\rho^{eq}} )  \} $  is the quantum relative entropy,\cite{vedral2002}   
and $\mathcal{F} = F + TS(\rho^{cc} || \rho^{eq})$ is the so-called information free energy.\cite{parrondo2015} In analogy with the perfect-Gibbs case, consider the non-Gibbsian infinitesimal of $dS_e$ :

\begin{equation}\begin{split}
dS_e& = \beta ( dE - d\mathcal{F}) = \beta \left( \Tr\{ (d\rho^{cc})\mathcal{H} \} + \Tr\{ \rho^{cc}(d\mathcal{H}) \}  - d\mathcal{F} \right) \\
\\
        & \equiv \beta ( \delta Q_{tot} - \delta Q_{cc} )
\bigskip
\end{split}\end{equation}

\noindent where $\delta Q_{tot} = Tr\{ (d\rho^{cc})\mathcal{H} \} $ is assimilated to the total heat exchanged, and $\delta Q_{cc} = d\mathcal{F} -\Tr\{ \rho^{cc}(d\mathcal{H}) \}$ is the "energetic price" to maintain coherence and correlation.

Then, the "entangled system" quantum-equivalent of the First Law can now be written as:

\begin{equation}
de = d S_e/\beta + \delta \mathcal{F} = Tr\{ (d\rho^{cc})\mathcal{H} \} + \Tr\{ \rho^{cc}(d\mathcal{H}) \}
\bigskip
\end{equation}

\noindent only formally similar to the previous statement, Eq.(\ref{eqweak}), but with the peculiarly different meaning of the symbols for $E, S_e, \mathcal{F}$.

\bigskip 

 \addcontentsline{toc}{subsection}{Quantum version of Landauer’s limit }

\noindent \textbf{Quantum version of Landauer’s limit. }
Consider a quantum system $S$ whose information content is progressively erased upon interacting with a quantum environment $E$.
Both $S$ and $E$ are living in their respective Hilbert spaces $\mathcal{W}_S,\mathcal{W}_E$.
Assume that the initial state of the composite system is factorized

\begin{equation}
\rho_{SE}(0) = \rho_S(0) \otimes \rho_E(0)
\bigskip
\end{equation}

\noindent such that no initial correlations are present. The environment is initially prepared in a thermal Gibbs state                                                             $\rho_E(0) = \exp(-\beta \mathcal{H}_E)/\mathcal{Z}_E$.
$S$ and $E$ interact via the unitary transformation $U(t)=\exp{(-i\mathcal{H}t/\hbar)}$, with  $\mathcal{H} = \mathcal{H}_S + \mathcal{H}_E + \mathcal{H}_{SE}$  the total Hamiltonian comprising the system, the environment and their interaction.

Landauer’s principle is related to the change in entropy of the total system plus environment, therefore we can reinterpret the heat exchanged 
between $S$ and $E$, as the difference between their respective initial and final entropy:

\begin{equation}
[ S(\rho^t_S) - S(\rho^0_S) ] + [ S(\rho^t_E) - S(\rho^0_E) ] = \Delta S_S - \Delta S_E = \mathcal{I}(\rho^t_{SE}) \ge 0
\bigskip
\end{equation}

This is, by definition, also equal to the \textit{quantum mutual information} exchanged between $S$ and $E$:

\begin{equation}
\mathcal{I}(\rho^t_{SE}) =  S(\rho^t_E)+S(\rho^t_E) - S(\rho^t_{SE})  
\label{qinfo}
\bigskip
\end{equation}

\noindent (Note that for a completely factorized initial state, $\mathcal{I}(\rho^t_{SE}) =0$.)
With some algebra (see full derivation in Ref.\cite{reeb14}) it is shown that the average heat dumped from  $S$  into the environment, $\langle Q_E \rangle = \Tr\{ (\rho^t_E - \rho^0_E) \mathcal{H}_E \}$, is equal to:

\begin{equation}
\beta \langle Q_E \rangle = \Delta S_S + \mathcal{I}(\rho^t_{SE}) + S(\rho^t_E || \rho^0_E)
\bigskip
\end{equation}

And since both  $\mathcal{I}(\rho^t_{SE})$ and $S(\rho^t_E || \rho^0_E) \ge 0$, it is also:

\begin{equation}
\beta \langle Q_E \rangle \ge \Delta S_S
\label{eqbeta}
\bigskip
\end{equation}

This important relationship therefore establishes that the only heat dissipation in quantum computing occurs during state initialization and reset (erasure) operations, which are both linear in the number of qubits: the entropy changes in the quantum system turn into heating of the environment, by an amount simply proportional to the number of qubits, and not to the dimension of their Hilbert space. That's quite good news, since for $N$ qubits the Hilbert space has dimension $2^N$ or, in other words, $2^N$-distinct possible eigenstates, a number that grows very quickly.
A classical computer simulating this quantum computer, instead, must use an energy at least equal to 2$^N k_B T \ln$2 just to initialize or erase the configuration. Hence, this represents an additional bound to quantum advantage for a given classical calculation.

\begin{figure}[h]
\centering
\includegraphics[width=0.9\textwidth]{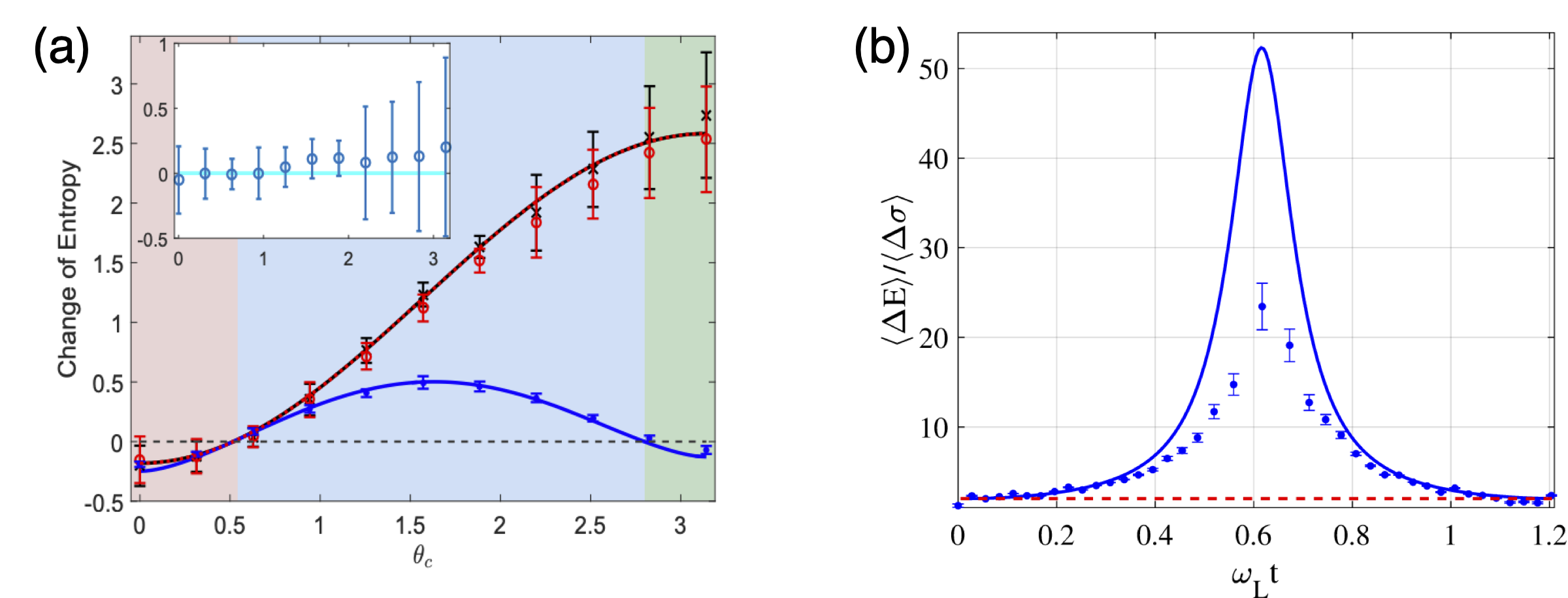}
\caption{ \textbf{(a)} Experimental verification of Eq.(\ref{eqbeta}) on a $^{40}$Ca ion-trap qubit. 
Black data $=\beta \langle Q_E \rangle$; red data $=\Delta S_S +  \mathcal{I}(\rho^t_{SE}) + S(\rho^t_E || \rho^0_E)$ (reprinted w. perm. from Ref.\cite{yan18}). 
\textbf{(b)} The $\beta \langle Q_E \rangle \ge \Delta S_S$  limit demonstrated experimentally  on a toy-model optical qubit gate. (Reprinted w/perm. from Ref.\cite{cimini20}). 
}
\label{figQB}
\end{figure}

Equation (\ref{eqbeta}) has been verified experimentally in a number of cases. In Figure \ref{figQB} the results of two such experiments are reported.\cite{yan18,cimini20}.

\bigskip

 \addcontentsline{toc}{subsection}{Thermalization: randomization of pure states into mixed states }

\noindent \textbf{Thermalization: randomization of pure states into mixed states.}
The typical initial condition of a quantum computer is a \textit{pure state}, for example with all the qubits prepared in a same state
$\ket{\psi_i} =\left(\begin{smallmatrix}1 \\ 0\end{smallmatrix}\right)$  for $i\in N$, by a previous RESET operation. As we saw in the subsection above, this operation costs both energy and heat, but it is fortunately linear with $N$.
Quantum decoherence explains how a system interacting with an environment, transitions from being a pure state (which exhibits coherent superpositions) to a mixed state, that is an incoherent combination of classical alternatives. 
The transition is ideally reversible, as the combined state of system and environment may still be a pure state. 
However, for all practical purposes it should be seen as irreversible, as the environment is in general a very large and complex quantum system, and it is not practically feasible to reverse their interaction. 

\smallskip

A  general description of the transformations between states when the quantum system is interacting with an external environment can be given by a kind of master equations, first introduced by Lindblad.\cite{lindblad} Such dynamics preserves trace and positivity of the density matrix, while allowing the density matrix to vary otherwise.\cite{petruccionebook} Master equations have the general form:\cite{manzano}

\begin{equation}
\frac{d\rho}{dt} = - \frac{i}{\hbar} \left[ \mathcal{H},\rho \right] + \sum_k \left[ L_k \rho L_k^{\dagger} - \frac{1}{2}\left( L_k^{\dagger}L_k\rho + \rho L_k L_k^{\dagger} \right) \right]
\label{lindblad}
\bigskip
\end{equation}

\begin{figure}[t]
\centering
\includegraphics[width=\textwidth]{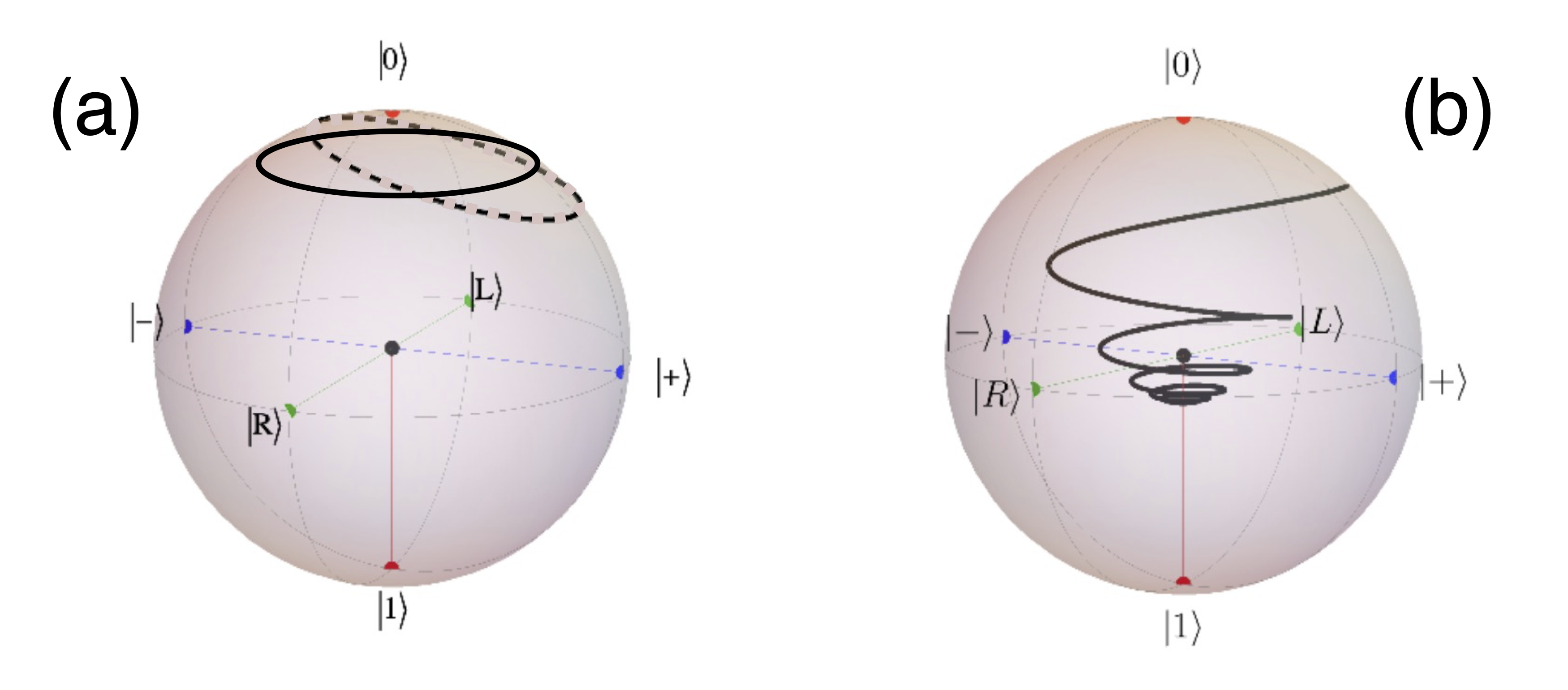}
\caption {\textbf{(a)} Unitary dynamics from Eq.(\ref{timevol}), and \textbf{(b)} dissipative dynamics from Eq.(\ref{master}) in the Bloch sphere. (Adapted from G.T. Landi, w/perm.)}
\label{lindbl}
\end{figure}

\noindent (to be compared with Eq.(\ref{timevol}) above). The $L_k$ are Lindblad operators that describe the effect of the interaction between the system
and the environment on the system’s state. A good example is the interaction of the 1/2-spin qubit with an electromagnetic field, for which there is just one operator $L=\sigma^+$, $L^{\dagger}=\sigma^-$, which applied on the qubit give $\sigma^-\ket{0}=\ket{1}$ and $\sigma^+\ket{1}=\ket{0}$. The external photon field is described by a spontaneous emission rate $\gamma_0$, with number density $N$ given by the Bose-Einstein distribution:

\begin{equation}
N = \frac{1}{e^{\beta\omega}-1}
\label{bosenst}
\bigskip
\end{equation}

\noindent and $\gamma=\gamma_0 (2N+1)$ is the total emission rate, including thermally-induced absorption and emission at the temperature $\beta=1/k_BT$.\cite{cherian19,morigi22}  The master equation describing the evolution is:

\begin{equation}
\frac{d\rho}{dt} = - \frac{i}{\hbar} \left[ \mathcal{H},\rho \right] + \gamma_0 (N+1) \mathcal{D}( \sigma^-) + \gamma_0 N  \mathcal{D}( \sigma^+ ) 
\label{master}
\bigskip
\end{equation}

\noindent where the more compact "dissipator" notation $\mathcal{D}(L)=L\rho L^{\dagger} - \tfrac{1}{2} \{L^{\dagger}L,\rho\} $ has been introduced. Figure \ref{lindbl}a,b  compares the evolution of  the density matrix $\rho(t)$ in the two cases: (a) under the action of Eq.(\ref{timevol}), with the unitary Hamiltonian $\mathcal{H}_0 =\hbar\omega \sigma_z/2$, and (b) the dissipative Eq.(\ref{master}). All the unitary Hamiltonian does is a precession of the state vector around $z$; on the other hand, upon coupling to the dissipator operator the precession is accompanied by a damping towards the $z$-axis. (Side note: The Hamiltonian for a spin population pumped by a coherent laser source, $\mathcal{H} = \mathcal{H}_0+\tfrac{\lambda}{2}(\sigma^+ + \sigma^-)$, is still unitary and hermitian (the rates of upward and downward transitions are equal), the result is just a precession about an axis inclined w/r to $z$, see dashed line in Fig.\ref{lindbl}a.)

\smallskip

Decoherence describes the classical limit of quantum mechanics, but is different from wavefunction collapse. 
In the mixed state  all  classical alternatives are still present, whereas the  wavefunction collapse (i.e., a measurement) 
selects only one of them.\cite{hill1997,wott1998}
Consider two qubits $A$ and $B$  (e.g., spin-1/2 particles, or polarized photons) characterized by an excitation energy $E$, at a temperature $k_BT=1/\beta$, so that the thermal probability of the excited state is $p=[1+\exp{(-\beta E)}]^{-1}$. Both a pure and a mixed state can be entangled over the ensemble of their qubits, however the difference between the two cases is important.
A pure state is called entangled when it is unfactorizable. The entanglement of a pure state can be defined as the (Von Neumann) entropy of either member of the pair. 
A mixed state, on the other hand, is called entangled if it cannot be represented as a mixture of factorizable pure states. The entanglement of a mixed state $\rho$ is the minimum average entanglement of an ensemble of pure states that represents $\rho$. 

The entanglement of formation $\epsilon_f(\rho)$  of the mixed state $\rho$, is the quantity of resources needed to create a given entangled state.\cite{bennett1996}  $\epsilon_f(\rho)$  is defined as the average entanglement between pure states of the decomposition, minimized over all the decompositions $\psi_i$ of $\rho$:\cite{vestra2001}

\begin{equation}
\epsilon_f(\rho) = \min \sum_i p_i S(\psi_i) = S_e \left[ \tfrac{1}{2} \left( 1 + \sqrt{1-C} \right) \right]
\bigskip
\end{equation}

\noindent also related to a different measure of entanglement, the \textit{concurrence} $C(\rho)$, via the Shannon entropy $S_e$.

In a quantum computation, maximally-entangled states of a pair of qubits (Bell states) can be constructed as we saw in Section 3 above, by applying a Hadamard gate (rotation) followed by a CNOT; let us call these two unitary operators $U_1$ and $U_2$. Starting from an initial density $\rho_i$, the final maximally entangled state is $\rho_f = U_2 U_1 \rho_i U_1^{\dagger} U_2^{\dagger}$.\cite{vestra2001}   What is interesting to note here, is that after some algebra, the concurrence of the final state can be obtained explicitly as:

\begin{equation}
C = \max \left( 0, 2p^2 - p - 2(p-1)\sqrt{p(1-p)} \right) 
\label{tempentg}
\bigskip
\end{equation}

This suggests the existence of an "entanglement threshold": for $p \lesssim 0.698$, or equivalently for $k_BT/E \gtrsim 1.19$, no entangled state of two qubits can be produced. For the typical TransMon excitation energy of the order of $E$=4 GHz, the maximum entanglement temperature is $T\simeq$240 mK. Note that this limit is well above the working temperature of SC loops, around 10-20 mK, while trapped ion qubits are operated at even lower, liquid-He temperatures.

\bigskip

 \addcontentsline{toc}{subsection}{Many qubits, multipartite systems }

\noindent \textbf{Many qubits, multipartite systems.}
We can associate to every unitary operation $U$ a work cost $W = \Tr{ \mathcal{H}(\rho^f-\rho^i)}$, which corresponds to the external energy input required to perform that operation. Think of two qubits $A$ and $B$ in a same thermal state at a temperature $T=1/\beta$; their initial thermal state is joint, we can write $\rho_{AB}(\beta)=\rho_A(\beta)\otimes\rho_B(\beta)$, but their Hamiltonian is non-interacting, $\mathcal{H}_{AB}=\mathcal{H}_A+\mathcal{H}_B$. To entangle (correlate) them we must bring the joint system out of equilibrium. Necessarily $W>0$ for every possible $U$ because the initial state is in thermal equilibrium. Then, a relevant question is: what is the minimal work cost for correlating thermal states? Or equivalently, what is the maximal amount of attainable correlations when the energy at our disposal is necessarily limited?

The result of a limiting temperature for the entanglement of a pair of qubits, obtained from Eq.(\ref{tempentg}), can be generalized to the case of multiple qubits.\cite{huber2015}  Consider $N$ qubits and the rotation Hadamard) from a pure to a maximally-entangled state in the subspace $\ket{0}^{\times N}$,  $\ket{1}^{\otimes N}$. Next, consider the possible bipartitions of the system $(j|N-j)$, for which we consider a subpart of the system of $j<N$ qubits entangled with its complement $N-j$. It has been shown that the concurrence is in fact independent on the particular choice of the bipartition, and depends only on the system size:

\begin{equation}
C = p^N - (1-p)^N -2p^{N/2} (1-p)^{N/2}
\bigskip
\end{equation}

Again, we ask what is the smallest thermal factor $p_b=[1+\exp{(-E/k_BT_b)}]^{-1}$, or the maximum temperature $k_B T_b=1/\beta_b$, which allows to simultaneously obtain entanglement across all the possible bipartitions of the system. By imposing $C$ to be positive, that is:

\begin{equation}
\frac{1}{\beta_b E} \geq  \frac{N}{2 \ln (1+\sqrt{2})}
\bigskip
\end{equation}

The corresponding work of correlation for the maximally-entangled set is:

\begin{equation}
W = NE \frac{(1-e^{-N \beta_b E})}{2(1+e^{-\beta_b E})^N} = NE \frac{1+\sqrt{2}}{\left[ (1+\sqrt{2})^{2/N} +1 \right]^N}
\bigskip
\end{equation}

\noindent which is exponentially small in the number of qubits $N$. This interesting result proves that by increasing the number of qubits, it becomes possible to generate partial entanglement even at (arbitrarily) high temperatures.
This is due to the fact that typical gate protocols act on qubit subspaces, whose population becomes negligible in the limit of large $N$. Therefore, even a small amount of entanglement obtained on a subset of the available states might be enough to obtain a substantial quantum advantage.  

\smallskip

How many correlations can be induced in a system of many qubits? And how can we make sure that a set of qubits is actually entangled? This is the more general problem of entanglement detection.\cite{guhne} A measure of the total number of correlations gives the deviation of the global state of the quantum computer from a corresponding uncorrelated state, a quantity that is important to estimate in the preparation of the initial correlated state. The total system composed of $k$ subsystems would be said to have zero correlation if its state is such that $\rho=\otimes \rho_i$, $i \in k$, i.e. the direct product of its partials. Therefore, a common measure of the correlation can be given by the relative entropy of the state:\cite{goold,horodeck,girolam}

\begin{equation}
S_{rel}(\rho) = \sum_i S_i(\rho_i) - S(\rho)
\bigskip
\end{equation}

Despite its apparent simplicity, such a measure is highly non-linear and difficult to access in a real experimental device with more than just a few qubits, so that alternative approaches have been proposed, based e.g. on the R\'enyi entropy,\cite{bridges} or the measurement of "witness" observables,\cite{guhne,friis}, or more general quantifiers including the notion of "fidelity".\cite{liang,zhenz}

\bigskip

\noindent \textbf{The supremacy clause of thermodynamics}. As it was briefly discussed in Section 4, the fundamental noise limit for classical computers is of thermal origin, with a contribution $\mathcal{O}(k_BT)$. To prevent random bit flipping, the excitation energy $E$ needs to be sufficiently larger than the thermal energy $k_BT$. Under reasonable assumptions, the error probability in assigning a state to a classical bit is:\cite{kish2002}

\begin{equation}
\epsilon_Q \gtrsim \exp{(-E/k_BT)}
\bigskip
\end{equation}

For a quantum computer with gates driven by auxiliary oscillators at frequency $\omega$ (with $\omega > 2\pi/\tau$, and $\tau$ the coherence time), the corresponding lower error limit is:\cite{bana2002}

\begin{equation}
\epsilon_Q \gtrsim \frac{\hbar \omega}{E} > \frac{h}{E\tau}
\label{eqerr}
\bigskip
\end{equation}

It can be noted that this is a sort of generalized time-energy uncertainty relation, describing the minimum energy needed to change a state in a time less than $\tau$ with failure probability smaller than $\epsilon_Q$. 
The meaning of this comparison is that in the quantum case the error decreases only in inverse proportion as the energy used, while in the classical case it decreases exponentially.
The quantum regime corresponds to $\hbar\omega > k_BT$, that is when the quantum noise in the driving oscillator exceeds the thermal noise at the work temperature.

\begin{figure}[h]
\centering
\includegraphics[width=0.75\textwidth]{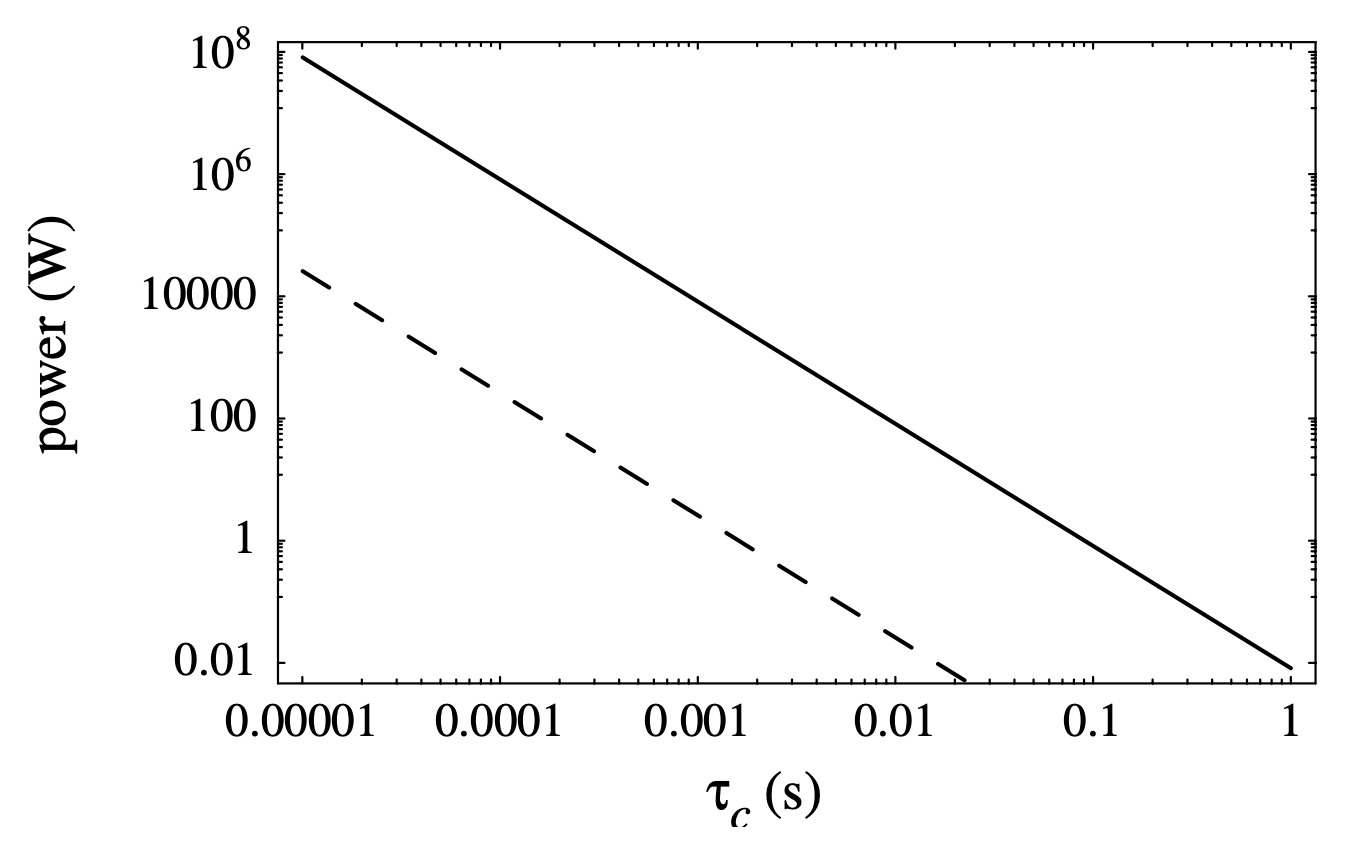}
\caption{Estimate of minimum power required to factor a 1000-bit number, as described in the text. Solid line: oscillatory control fields, 
$\omega=(\epsilon_Q^{3/2} \tau)^{-1}$. Dashed line: static control fields only. (Reprinted w/perm. from Ref.\cite{bana2002b}) }
\label{figpow}
\end{figure}

Because of quantum reversibility the excitation energy $E$ need not be dissipated, it only needs to be put into, 
and removed from, the driving system to switch on and off the desired gate evolution. In theory, if nothing is ever
erased, ‘‘conservative’’ computation is possible. However, there are two caveats.
Firstly, any proposed quantum computer architecture up to date has no mechanisms to actually “recycle” that energy. Current SC gates are based on microwave power pulses, which obviously consume energy irreversibly. However, any practical computer, even if it could use fully-reversible gates, is still going to generate heat at least because of error correction to keep the computation on track. Error correction inherently requires irreversible operations, such as supplying continuously ground-state configurations (e.g. "zero" ancilla qubits). Reversible circuits need to be adiabatic, there cannot be heat exchanges between the circuit and its environment. They must be in equilibrium at all times, which means to conserve the squared modulus of the wavefunction, and "all times" actually means during the coherence time.

Two universal reversible logic gates, both operating on three bits or qubits, have been suggested to implement logically-reversible operations. The Toffoli gate  \cite{toffoli}  inverts the state of a target bit conditioned on the state of two control bits; the Fredkin gate \cite{fredkin} swaps the last two bits conditioned on the state of the control bit. In these two gates, any set of inputs is processed and results as a unique pattern of outputs; these gates are therefore logically reversible. Examples of logically reversible circuits have actually been designed and fabricated,\cite{shao07,patel16,zhou22,orbach12} however they always practically display some degrees of energy dissipation by different means (e.g., requiring ancilla qubits,\cite{ikon17} finite-rate operation and read-out,\cite{orbach12} and so on).

Secondly, and most important, this much excitation energy to prepare the initial state, even if in principle recyclable, has to be fully available at least to start the calculation. If we assume as a practical upper bound 
 $\epsilon_Q\simeq 10^{-5}$ and a GHz quantum computer, the minimum (reversible) $E$ per elementary logical operation is 
then $\sim$0.007 eV, compared to the thermal energy $\sim 10^{-6}$ eV at 30 mK.
As a worst-case example,\cite{bana2002b} let us suppose that 5000 logical qubits are needed to factor a 1000-bit number; a 7-qubit code concatenated only once (depth=2) is used for error correction; only local gates are available; and about 10 ancilla qubits per logical qubit are used. 
Figure \ref{figpow} shows estimates of the minimum power as a function of coherence time, for a driving microwave field at frequency $\omega = (\epsilon_Q^{3/2} \tau)^{-1}$, the full and dashed lines corresponding to periodic and static excitation fields.

Despite the purely heuristic nature of Eq.(\ref{eqerr}) (for example, the error limit could be improved by smarter correction algorithms, or by improved hardware solutions) the results clearly indicate that, for very large-scale quantum computations, one really needs to use quantum systems
with very long decoherence times. Values of $\tau$ in the 100-$\mu$s range would require megawatt start-up power. It is just not feasible to get around the problem of short decoherence times just by driving the system at faster frequencies. This also suggests that there could probably never be the equivalent of a ‘‘Moore’s law’’ for quantum computers.

\bigskip

 \addcontentsline{toc}{subsection}{Objectivity of measurement and "Quantum Darwinism" }

\noindent \textbf{Objectivity of measurement and "Quantum Darwinism"}
In the standard circuit (or QED) model, the array of qubits is initialized for example in the logical $\ket{0}$ state; then, a sequence
of quantum gates is applied depending on the required algorithm; finally, a read-out operation is
carried out by measuring individual qubits in the same $\ket{0/1}$ computational basis.  In alternative, the adiabatic quantum computation does not rely on gate sequences, but on the direct implementation of a smoothly varying Hamiltonian on the network of qubits; after the initial prepared state, annealing and read-out are cyclically performed to obtain the global optimum configuration of spins, which gives the ground state of the "solution" Hamiltonian. In either instance, the read-out operations give a human-readable, classical physics result from the quantum computation.

The final state of the computation is something like $\ket{\psi}=\sum_n c_n \ket{n}$ and, as we know, the complex amplitudes $c_n$ are not directly accessible. The measurement gives one of the $n \in N$ possible outcomes with probability $|c_n|^2$, that is a probabilistic rule for projecting the state vector onto one of the vectors of the orthonormal measurement basis. Consider for example the general qubit state $\ket{\psi}=c_1 \ket{0} + c_2 \ket{1}$, and assume that we want to perform a measurement in the orthonormal basis $\ket{u}=a\ket{0}+b\ket{1}$, $\ket{v}=b^*\ket{0}-a^*\ket{1}$. The probability of a measurement giving $\ket{u}$ as a result is:

\begin{equation}
P(u) = |\braket{u,\psi}|^2 = \\
         | (a^*\bra{0} + b^*\bra{1}) (c_1 \ket{0} + c_2 \ket{1})|^2 = \\
         | a^*c_1 + b^*c_2 |^2
\bigskip
\end{equation}

\noindent and similarly, the probability of getting $\ket{v}$: 

\begin{equation}
P(v) = |\braket{v,\psi}|^2 = \\
         | (b\bra{0} - a\bra{1}) (c_1 \ket{0} + c_2 \ket{1})|^2 = \\
         | bc_1 - ac_2 |^2
\bigskip
\end{equation}

Decoherence of the qubits is the loss of their typical quantum properties, entanglement and non-locality, through interactions with the environment. New correlations with the thermal bath degrees of freedom appear, which degrade the information originally encoded in the quantum system.

In classical physics, what you see is simply “how things are”. You can measure a tennis ball traveling at 120 km/h to a given direction, passing through a given point in space at a given instant of time. What more is there to say?
But when a quantum particle is in a state of “superposition” before the measurement, the various superposed states interfere with one another in a wavelike manner. Only when we make a measurement we see one of those outcomes. But, given the probabilistic nature of the result, why just \textit{that one}? Could someone else check our result and find that same outcome?

The definite properties that we associate with classical physics, such as position and velocity, may be selected from a “menu” of quantum possibilities, in a process loosely analogous to natural selection in evolution. The quantum properties that survive are - in a kind of pseudo-Darwinist sense - the "fittest".\cite{zurek1982,zurek2003}  
And, as it happens in natural selection, the “survivors” are those that make the most copies of themselves. Many independent observers can thus make measurements of the quantum system, each one using a different copy of the result, and agree on the outcome - a hallmark of classical behavior. "Quantum Darwinism" (QD, \cite{zwolak,palma,pan19,pater21}) changes the role of the environment, from being a shady background with undetermined characteristics, into a fragmented space filled with redundant information that can be accessed and "measured" by individual observers. Notably, this notion is different from the macroscopic limit, because of which the number of degrees of freedom of the environment is so large that the \textit{averaging} process dominates the read-out process; QD instead deals with the mechanism by which the quantum information gets encoded (i.e., entangled) to the surrounding quantum states of the environment.

Experimental measurements of the result of a quantum computation are typically recorded by collecting information transmitted through some carriers - photons, electrons, phonons - that constitute the environment (thermal bath).
While there will be many such individual information carriers, only a small fraction typically needs to be captured in order for the observer to accurately record the measurement. 
Given two observers, they will agree on the outcome when they can independently intercept different fractions of these information carriers, and both perform the same type of measurement on their respective sets. In the QD scheme,
they will necessarily arrive at the same conclusion, due to the entanglement shared between the system and all the environmental degrees of freedom. 
Then, a key question is whether is it possible to get enough information, by monitoring only a small part of the environment?

\begin{figure}[t]
\centering
\includegraphics[width=0.85\textwidth]{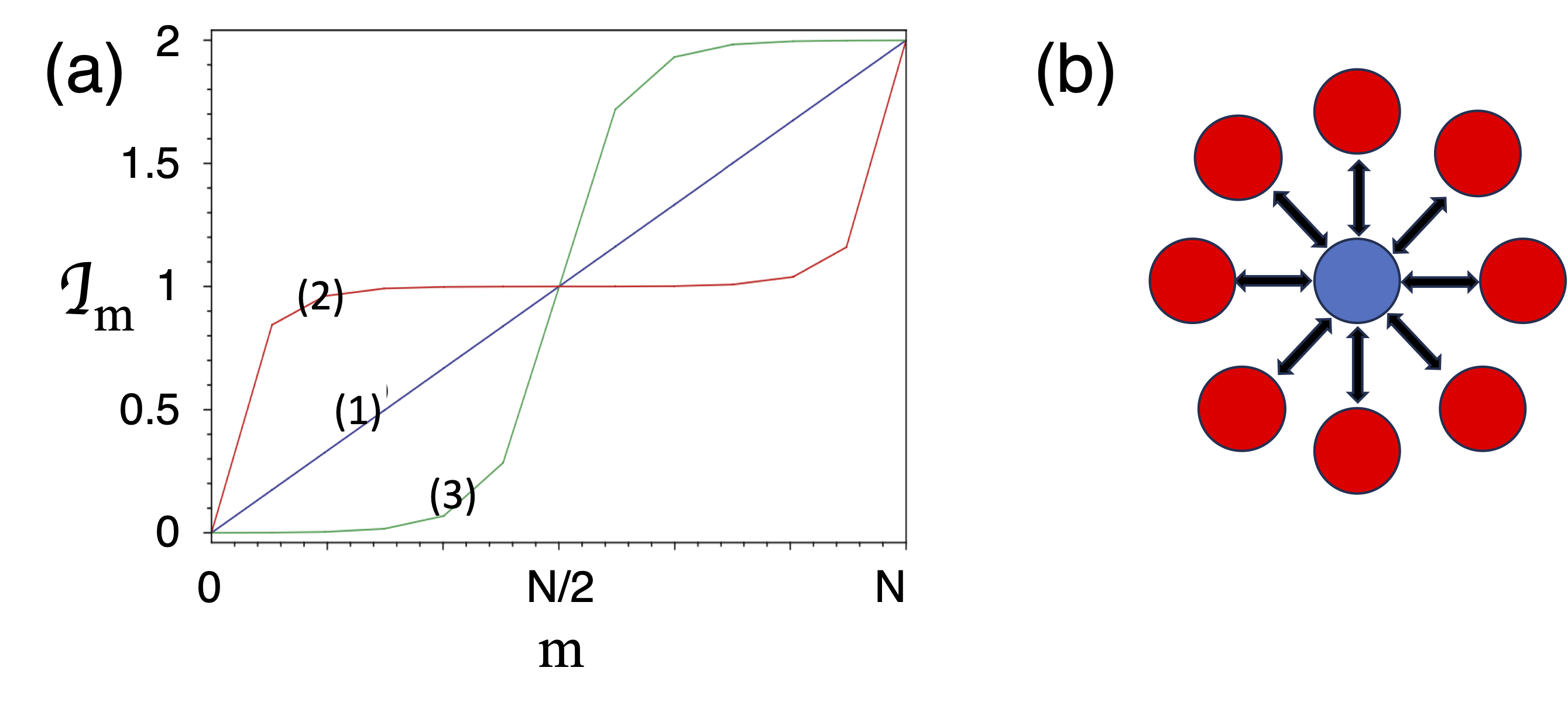}
\caption{\textbf{(a)} Typical behavior of the information plots $\mathcal{I}_m$ as a function of the fractions of environment interrogated by different observers. \textbf{(b)} The spin star environment. }
\label{darwin}
\end{figure}

We may look at the amount of (Shannon) entropy that is produced by destroying the correlations between the system $S$ and a fraction $m \in N$ of the total environment $E$, that is the quantum mutual information $\mathcal{I}$ defined in Eq.(\ref{qinfo}) above, and ask how the partial information gathered compares to the whole.\cite{zurek05} From the obvious condition that $\mathcal{I}_m$ must be non-decreasing, three possible behaviors can be envisaged as shown in Figure \ref{darwin}a: the linear one, $\mathcal{I}_m \propto m$, in which each fraction of the environment provides unique and independent information, so that each observer would obtain a separate information about the system: alternatively the curve 2, describing redundantly stored information, $\mathcal{I}_m$ rapidly increases, then plateaus at the value for which all observers essentially obtain the same information (the so-called "objectivity plateau"); or the curve 3, describing information about the system that is tightly encoded, so that $\mathcal{I}_m$ remains close to zero, then suddenly increases to the maximum around some characteristic amount, for example $m \sim N/2$. 

\smallskip 

Let us consider a quantum system $S$  of a single qubit initially in a pure state, superposition of two states $\ket{\psi_1}, \ket{\psi_2}$  expressed in the conventional basis $\ket{0/1}$ (for the sake of simplicity, I avoid here the customary introduction of the "pointer" states):

\begin{equation}
\ket{\psi_S} = a \ket{\psi_1} + b \ket{\psi_2}, \,\,\,\,\,\,\,\,\,\,\,\, |a|^2 + |b|^2 = 1
\bigskip
\end{equation}

\noindent and embedded in an “environment” of $N$ other qubits, all in a same generic state $\ket{\psi_E}$,  also expressed in the same basis $\ket{0}, \ket{1}$ but with random coefficients.  The overall initial state is assumed to be without correlation between $S$ and $E$, i.e. factorizable as :

\begin{equation}
\ket{\psi^0_{SE}} = \left( a \ket{\psi_1} + b \ket{\psi_2} \right) \otimes \ket{\psi_E}
\bigskip
\end{equation}

QD posits that, after some thermalization time, in which the coupled $S-E$ system evolves, the total state of the system can be turned to:

\begin{equation}
\ket{\psi^t_{SE}} =  a \ket{1}\ket{1^{\otimes N}} + b \ket{0} \ket{0^{\otimes N}}
\bigskip
\end{equation}

\noindent the notation $M^{\otimes N}=M\otimes M ... \otimes M$ indicates the direct product of all environment degrees being in either one or the other state of the computational basis.

Several authors have considered the configuration as a "spin star" (e.g., \cite{,giorgi15,pater21}, see Fig.\ref{darwin}b), in which the single qubit is a spin surrounded by a circle of environmental spins. Different subgroups of environment spins can be read-out by different observers, without perturbing the central spin, which interacts independently and equally with each one of the subsystems


If now we take the partial trace over the $N$ qubits (spins) of the environment, the density matrix for $S$ is obtained:

\begin{equation}
\rho_S =  |a|^2 \ket{\psi_1}\bra{\psi_1} +|b|^2 \ket{\psi_2} \bra{\psi_2}
\bigskip
\end{equation}

\noindent while each of the environment qubits has the same density matrix:

\begin{equation}
\rho_{E_i} =  |a|^2 \ket{0}\bra{0} +|b|^2 \ket{1} \bra{1}
\bigskip
\end{equation}

The crucial point is that, although the system $S$ has lost its coherence, the population coefficients $(a,b)$ of the qubit $S$ are "imprinted" on each of the $N$ environment qubits, generating a redundancy of information. This is the phenomenon manifested in the "plateau" of constant information seen in Fig.\ref{darwin}a, curve (2). Then, different observers measuring only a subset of the final state will agree on the result. And this should represent the emergence of classical objectivity.

\section{Conclusions}\label{sec13}

This overview tried to provide a (necessarily limited and incomplete) synthesis of some outstanding issues in the definition of thermodynamic concepts at the level of quantum mechanics, under the peculiar angle of their possible and likely impact on quantum computer technology, and quantum computing algorithms. This field has known a rapid growth in the past decades, moving from the domain of theoretical speculations, to the urgent requirement of starting to provide real solutions to practical problems that the quantum computing hardware is facing. Despite the main technical difficulties today still lie in the probabilistic nature of the quantum computing output and the need for error correction, it is possible that thermal limits will represent the next hurdle for the efficient and useful operation of such machines. 

It may look surprising that the historical and philosophical discussions about the Second Law of thermodynamics should have an interest, and even represent a foundation for practical quantum computing. The relationship between information theory, manipulation of information at small scales (which also interests other fields, such as molecular and DNA-based computing \cite{kempes,daley}) and thermodynamics is not purely formal, but treats information as a physical entity. The contribution of fluctuation theorems and stochastic thermodynamics provides a more ample framework for analyzing quantum information and exchanges of work and heat in open quantum systems. The definition of quantum entropy (Von Neumann's, despite some ambiguities, or other competing definitions) is also key in the attempt at understanding the emergence of macroscopic information in the measurement process.

Still, several problems and questions remain open, both at the fundamental- and applied-physics level. For example, the definition of quantum equivalents of work and heat given in Eq.(\ref{eqweak}) and the path-integral form in Eq.(\ref{pathw}) refer to different situations. While the latter, fluctuation-based concept is applicable in general to either closed or open systems, the "weak" form refers to the average energy exchanges (ensemble averages) in and out of the system. In most cases these (and a couple other) different definitions arrive in practice at the same results, however our understanding of the question still appears not solid enough, and open to further investigation. 

At first, entanglement seems to be unrelated to thermodynamics. However, the challenge of maintaining entangled states is linked to the interaction of qubits with the environment, that is a thermal bath. Quantum decoherence, or the loss of off-diagonal components in the density matrix, is the process that eventually undermines entanglement, by transfer of entangled states between the computing qubits and the environment's quantum states. There is a whole thermodynamic domain that I did not touch in this article, that is quantum batteries,\cite{dutta21,batt22} whose key problem is to quantify the maximum extractable work, and which crucially depends on the interplay of coherence and entanglement between the quantum battery and the charger.

Quantum computers can check and verify the theoretical predictions of quantum thermodynamics, and quantum thermodynamics will, in turn, help to quantify and master dissipative processes in quantum computing. The interplay of quantum mechanics and thermodynamics is a young research field, still rich of interesting issues and open questions.

\backmatter

\bmhead{Acknowledgements}

I gratefully thank my colleagues Val\'erie Vallet and Stephan  De Bievre for their kind invitation to the Quantum Information Working Group, in the University of Lille, which provided the excuse to assemble these lectures.

\bmhead{Competing interests}

The author declares no competing interests.

\bmhead{Funding}

Institutional funding from IEMN CNRS and the University of Lille is acknowledged. 

\bmhead{Data availability}

All data that is used is public and referred to by references, footnotes or otherwise in the article.

\bmhead{Author's information}

Fabrizio Cleri is Distinguished Professor of Physics in the Lille University, France, and director of the Physics Division at the local CNRS Institute of Electronics, Microelectronics and Nanotechnology (IEMN UMR 8520). He graduated \textit{cum laude} in theoretical nuclear physics at the University of Perugia, Italy (1985), and received the \textit{Habilitation} in physics and materials science at the University of Strasbourg, France (2005). 

 \addcontentsline{toc}{section}{References }

\bibliography{sn-bibliography}

\end{document}